\documentclass[aps,12pt,nobibnotes,nofootinbib,longbibliography, a4paper,superscriptaddress,notitlepage,citeautoscript]{revtex4-1}  

\usepackage{amsmath}
\usepackage{mathrsfs}
\usepackage{amssymb}
\usepackage{graphicx,epstopdf,epsfig}
\usepackage{bm}
\usepackage{color}
\usepackage{times}
\usepackage{hyperref}

\usepackage{bm}
\newcommand{\mr}[1]{{\mathrm{#1}}} 			
\newcommand{\mrb}[1]{\bm{\mathrm{#1}}} 		
\newcommand{\ii}{{\rm i}}						
\newcommand{\ket}[1]{{\vert #1 \rangle}} 			
\newcommand{\Omp}{\,\Omega_{\rm p}}			
\newcommand{\Omc}{\,\Omega_{\rm c}}			
\newcommand{\Ge}{\,\Gamma_e}				
\newcommand{\e}{{\rm e}}						
\newcommand{\N}{{N_{\rm b}}} 				

\newcommand{\eref}[1]{Eq.~(\ref{#1})}
\newcommand{\sref}[1]{Sec.~\ref{#1}}
\newcommand{\cref}[1]{Chap.~\ref{#1}}
\newcommand{\fref}[1]{Fig.~\ref{#1}}

\newcommand{\Cref}[1]{Chapter~\ref{#1}}
\newcommand{\Fref}[1]{Figure~\ref{#1}}

\newcommand{\etal}{\textit{et al.}}

\begin{document}
\title{Non-linear optics using cold Rydberg atoms}

\author{Jonathan D. Pritchard}
\affiliation{Department of Physics, University of Strathclyde, Glasgow, G4 0NG}
\author{Kevin J. Weatherill}
\author{Charles S. Adams}
\email{c.s.adams@durham.ac.uk}
\affiliation{ Department of Physics, Durham University, Durham, DH1 3LE}

\begin{abstract}
The implementation of electromagnetically induced transparency (EIT) in a cold Rydberg gas provides an attractive route towards strong photon--photon interactions and fully deterministic all-optical quantum information processing. In this brief review we discuss the underlying principles of how large single photon non-linearities are achieved in this system and describe experimental progress to date.
\end{abstract}
\maketitle

\tableofcontents



\section{Introduction}

Non-linear optics arises when the interaction between light and matter depends on the intensity of the applied electromagnetic fields  \cite{boyd08}. The non-linear effect may involve only one field (as in self phase modulation or frequency doubling) or many fields (as in cross phase modulation or four-wave mixing). A non-linear response appears when the optical resonance frequency depends on the intensity of a control field for example via the AC Stark effect. The non-linearity is typically proportional to the magnitude of the resonant shift; consequently states with large sensitivity to external fields have enhanced non-linearities.

In this review we focus on the case of highly excited Rydberg states where one electron is excited into a state with a large average separation from the nucleus. For a Bohr--like atom with the outer electron in a state with principle quantum number $n$, the sensitivity to low frequency electric fields scales as $n^7$ and to other Rydberg atoms as $n^{11}$. Therefore by exploiting states with higher principal quantum numbers one can achieve an enormous enhancement in the sensitivity to electric fields and other Rydberg excitations. Below we will highlight two obvious applications of this enhanced sensitivity, first in electrometry, and secondly in achieving strong photon--photon interactions which could have enormous implications for quantum information processing using light.

Photons are ideal carriers of classical or quantum information due to their weak free-space interactions. The flip side of these weak interactions is that photons are relatively difficult to control at the `single quantum' level. Consequently, most single photon processes depend on probabilistic protocols, for example linear-optic-quantum-computing (LOQC)  \cite{knill01,duan01} with only a small probability of success. Using Rydberg non-linear optics it is possible to achieve strong photon--photon interactions by mapping the photon onto Rydberg excitations which are coupled via long-range dipole--dipole interactions. This gives rise to a large optical non-linearity at the single photon level and therefore {the potential to realise fully deterministic protocols for manipulating photons.}

The review is organized as follows: First we review the basic concepts of non-linear optics and then in \sref{sec:Rydberg} we introduce the principle of Rydberg non-linear optics. \sref{sec:expt} contains a review of experiments to date and finally in \sref{sec:qo} we discuss extending Rydberg non-linear optics into the quantum domain.

\subsection{Optical Kerr non-linearities}
To understand why optical non-linearities are small, and hence why photon--photon interactions are extremely weak, we begin with a brief classical outline of the origin of the non-linear optical response of atoms: as light propagates through an atomic medium, the electric field ${\cal E}$ of the electromagnetic wave induces an average dipole $\langle d\rangle$ per atom. This additional dipolar field results in attenuation and a phase-shift of the total field. These effects are characterized by the real and imaginary components of the {electric} susceptibility, $\chi$, which for a medium with {a density of} $N$ dipoles per unit volume is given by
\begin{eqnarray}
\chi&=&\frac{N\langle d\rangle}{\epsilon_0 {\cal E}}~.
\label{eq:chi}
\end{eqnarray}
Non-linearities arise when the induced dipole $\langle d\rangle$ is a non-linear function of the incident field ${\cal E}$. Classically, this occurs due to the anharmonic response of the oscillating charge. In this case we expand the susceptibility as a power series in ${\cal E}$
\begin{eqnarray}
\chi&=&\frac{N\langle d\rangle}{\epsilon_0 {\cal E}}
=\chi^{(1)}+\chi^{(2)}{\cal E}+\chi^{(3)}{\cal E}^2+\cdots~,
\label{eq:chi_nl}
\end{eqnarray}
where $\chi^{(1)}$ is the linear response and higher order terms quantify the non-linear response.

\begin{figure}[!t]
\begin{center}
\includegraphics[width=7.5cm,angle=0]{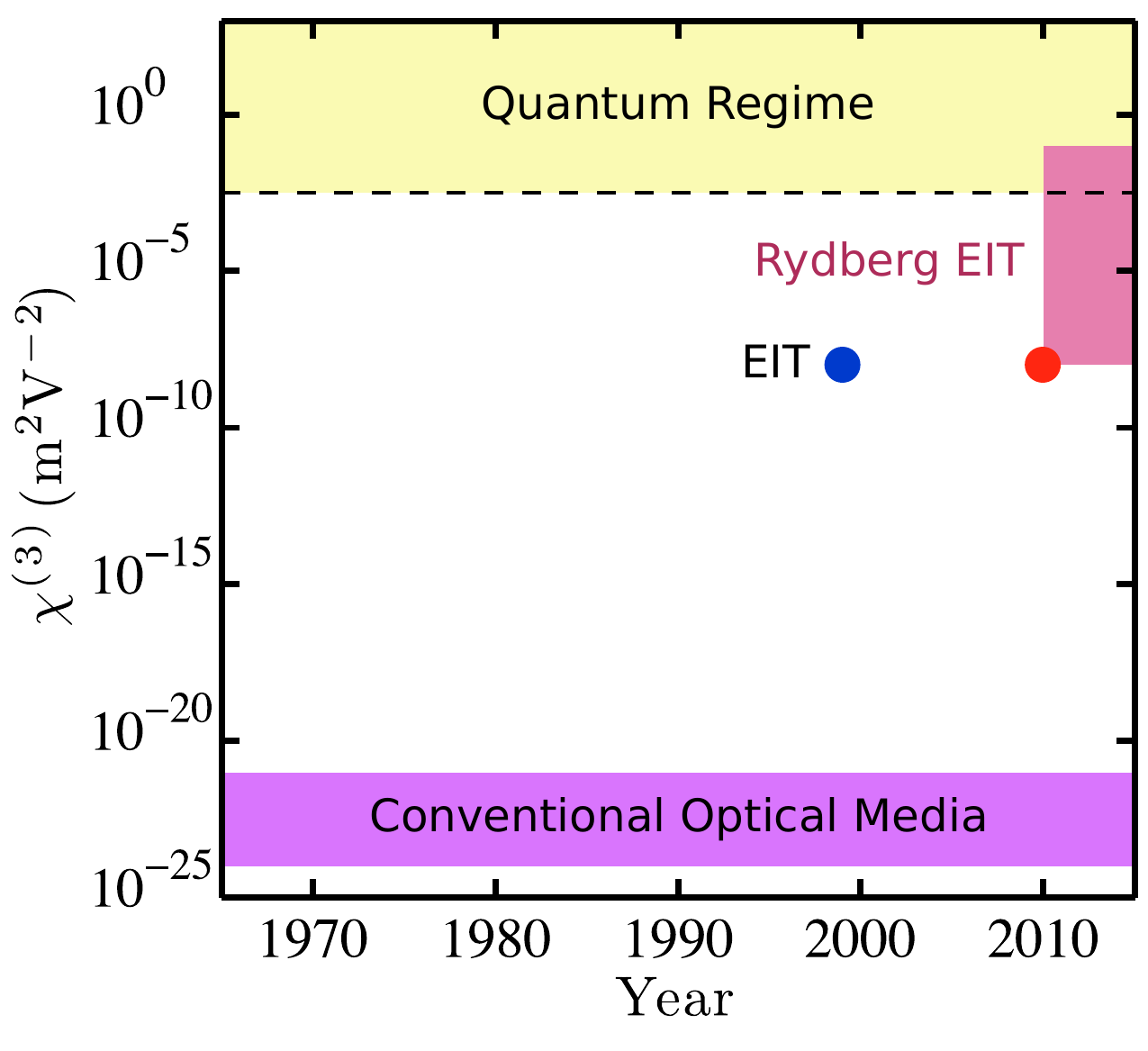}\caption[]{Schematic showing advances in the optical Kerr non--linear coefficient, $\chi^{(3)}$. Conventional optical materials have resonances in the ultra-violet leading to a small non-linearity in the visible and infra-red. Enormous enhancements of the non-linearity are possible by using resonant media but at the expense of loss. The loss can be reduced using the technique of electromagnetically induced transparency (EIT). The giant EIT non-linearity can be further enhanced using a Rydberg medium with strong dipole--dipole interaction. With Rydberg EIT it is possible to exceed the  threshold for single photon non-linearities (dashed line) where a quantum description is required. Experimental measurements are indicated by the dots. The EIT experiment corresponds to L. V. Hau \etal{} \cite{hau99} and Rydberg EIT to J. D. Pritchard \etal{} \cite{prit10}. The shaded area labeled Rydberg EIT indicates the extrapolation to higher density \cite{sevi11b}.
}
\label{fig:chi3}
\end{center}
\end{figure}

The second-order non-linearity $\chi^{(2)}$ vanishes for optical media possessing inversion symmetry; thus for atoms the dominant non-linear term is $\chi^{(3)}$. This is known as the optical Kerr non-linearity, which gives rise to cross-- and self--phase modulation  \cite{boyd08}. In very simple terms a field ${\cal E}$ induces a phase shift of order
\begin{eqnarray}\label{eq:phi}
\Delta \phi&=&k\ell\chi_{\rm r}^{(3)}\mathcal{E}^2~,
\end{eqnarray}
where the subscript r denotes the real component of the susceptibility, $k=2\pi/\lambda$ is the wavevector of the field and $\ell$ is the length of the medium. A deterministic photonic phase gate would require that this phase shift is of order $\pi$ when the field $\mathcal{E}$ is due to a single photon. To estimate the single photon phase shift we first write that the non-linearity is proportional to the ratio of the applied field to the binding field of electrons inside the atom \cite{boyd08}
\begin{eqnarray}
\chi_{\rm r}^{(3)}{\cal E}^2=\chi_{\rm r}^{(1)}\frac{{\cal E}^2}{{\cal E}_{\rm at}^2}~.
\end{eqnarray}
For a Bohr atom ${\cal E}_{\rm at}\sim\textstyle{1\over 2}  e/4\pi \epsilon_0 a_0^2\sim 5\times 10^{11}$~Vm$^{-1}$, and as $\chi^{(1)}\leq 1$\footnote{
The magnitude of the linear susceptibility $\chi_0=\vert\chi^{(1)}\vert=6\pi Nk^{-3}$, where $Nk^{-3}$ is the number density multiplied by the reduced wavelength cubed which gives the number of atoms in a cube with sides of length $\lambda/2\pi$. For typical densities in cold atomic ensembles  $6\pi Nk^{-3}\ll 1$, even for the high densities achieved in a Bose-Einstein Condensate (BEC) e.g. $N=10^{14}$~cm$^{-3}$ in Rb, $\chi_0=6\pi Nk^{-3}=0.36$.
} we obtain
\begin{eqnarray}
\left\vert \chi_{\rm r}^{(3)}\right\vert &\leq& 10^{-23}~{\rm V}^{-2}{\rm m}^2~.
\end{eqnarray}
This rough estimate agrees well with typical values of conventional optical media, for example air is $1.7\times 10^{-25}~{\rm V}^{-2}{\rm m}^2$ and water is $2.5\times 10^{-22}~{\rm V}^{-2}{\rm m}^2$ \cite{boyd08}.  This range of typical optical non-linearities is illustrated in \fref{fig:chi3}.

To estimate the electric field of a single photon field we assume a bandwidth of 1 MHz (which determines the length of the single photon wavepacket) and a beam size of order of the wavelength which gives ${\cal E}\sim 10~{\rm V}{\rm m}^{-1}$.\footnote{We use
\begin{eqnarray}\label{eq:photon}
{\cal E} \sim \left(\frac{\hbar\omega \Delta \omega}{2\pi^2\epsilon_0 w_0^2}\right)^2~,
\end{eqnarray}
where $\Delta\omega$ is the bandwidth and $w_0$ is the beam radius. This relation ignores diffraction so would apply to, for example, light confined in a hollow core fiber.
}
Consequently optical non-linearities in conventional media are about 20 orders of magnitude smaller than that required for single photon non-linear optics.

A much larger non-linearity can be obtained by exploiting a resonant transition in the medium. For the case of a two-level atom with states $\ket{g}$ and $\ket{e}$ driven by an optical field detuned by $\Delta=\omega-\omega_0$ from the resonant frequency $\omega_0$, the susceptibility is given by
\begin{eqnarray}\label{eq:chi2}
\chi= \chi_0\frac{\ii\Gamma/2}{\Gamma/2-\ii\Delta}~,
\end{eqnarray}
where the transition between states $|g\rangle$ and $|e\rangle$ has linewidth $\Gamma$ which is related to the dipole moment $d_{\rm ge}$ by $\Gamma=k^3d_{\rm ge}^2/(3\pi\epsilon_0\hbar)$. The electric field of the light causes an AC Stark shift of the resonant frequency of the atomic dipole, leading to $\Delta\rightarrow\Delta-\alpha\mathcal{E}^2/(2\hbar)$. Inserting this into \eref{eq:chi2} and using the off-resonant polarizability $\alpha = 2d_{\rm ge}^2/(\hbar\omega_0)$, the resulting Kerr non-linearity is
\begin{eqnarray}
\chi_{\rm r}^{(3)}= \chi_{\rm r}^{(1)}\frac{d_{\rm ge}^2}{(\hbar\omega_0)^2}~,
\end{eqnarray}
valid assuming that the detuning $\vert \Delta\vert $ is larger than the transition linewidth $\Gamma$. For smaller detunings $\chi^{(3)}$ is resonantly enhanced, however this coincides with the maximum scattering rate due to the imaginary component of the susceptibility which attenuates the transmission as $\exp(-k\chi_{\rm I}\ell)$. In other words the largest non-linearity occurs when the light transmission is a minimum.

\subsection{Electromagnetically Induced Transparency}\label{sec:EIT}
\begin{figure}[!t]
\begin{center}
\includegraphics{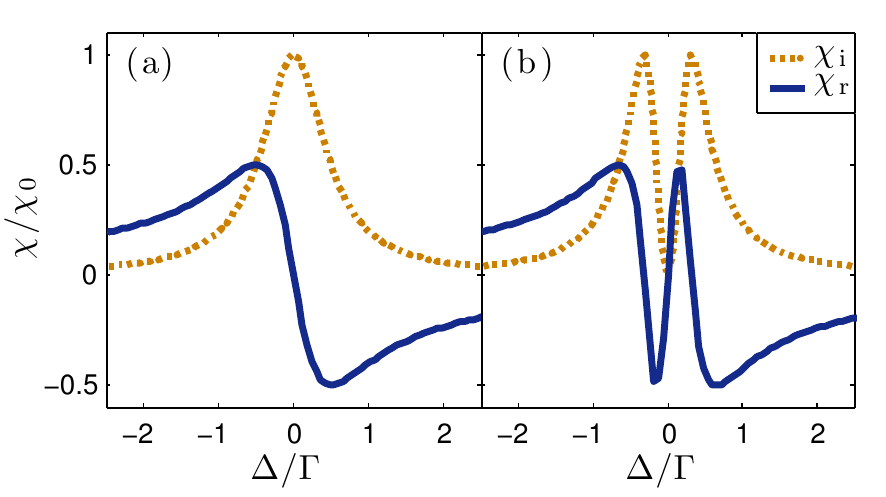}
\caption[]{The real ($\chi_{\rm r}$) and imaginary ($\chi_{\rm i}$) parts of the susceptibility for (a) an ensemble of two-level dipoles and (b)  for electromagnetically induced transparency (EIT) in an ensemble of three level atoms. The $\chi_{\rm r}$ and $\chi_{\rm i}$ determine the off-axis scattering (absorption) and the phase shift (refractive index), respectively. }
\label{fig:chi}
\end{center}
\end{figure}

In the 1990s it was realized that we can circumvent this problem by adding a strong resonant coupling from the intermediate state $\ket{e}$ to an additional level $\ket{r}$ to cancel the absorption on resonance. This is know as electromagnetically induced transparency (EIT) \cite{harris90,boller91}. There are excellent theoretical treatments of EIT in Gea-Banacloche {\it et al}.  \cite{geabanacloche95} and the review by Fleischhauer {\it et al} \cite{fleischhauer05}, who derive the susceptibility at the probe laser frequency as
\begin{eqnarray}\label{eq:chieit}
\chi= \chi_0\frac{\ii\Gamma/2}{\Gamma/2-\ii\Delta+\frac{\Omc^2/4}{\gamma_{\rm gr}-\ii\Delta}}~.
\end{eqnarray}
Here $\Omc$ is the Rabi frequency of the strong coupling laser and $\gamma_{\rm gr}$ is the dephasing rate of the coherence between $\ket{g}$ and $\ket{r}$. Typically $\gamma_{\rm gr}\ll\Gamma$, resulting in a narrow transmission window appearing at the center of the two-level absorption feature. This can be seen from \fref{fig:chi} which plots the real and imaginary components of the susceptibility as a function of $\Delta$ for EIT compared to that of a two-level atom. Associated with the vanishing absorption on resonance, the real component of the susceptibility reveals a steep dispersive feature corresponding to a large group velocity, and hence slow light.

On resonance one can now achieve a Kerr non-linearity of order
\begin{eqnarray}
\chi_{\rm r}^{(3)} \approx \chi_{\rm r}^{(1)}
\frac{d_{\rm ge}^2}{(\hbar\Omega_{\rm c})^2}~.
\end{eqnarray}
The enormous enhancement compared to the off--resonant case is immediately apparent as we have replaced photon energy $\hbar\omega_0$ by $\hbar\Omega_{\rm c}$ which differ by 8 orders of magnitude.

The first experiments demonstrating such large non-linearities in an EIT medium were performed by L. V. Hau \etal{} \cite{hau99} who slowed light to 17~ms$^{-1}$ in a Na ensemble close to the BEC transition. The measured Kerr coefficient was $\chi_{\rm r}^{(3)}\sim 7\times 10^{-8}~{\rm V}^{-2}{\rm m}^2$, one of largest Kerr like non--linearities ever reported but still too small for single photon non--linear optics. Using conventional non-linear optics it is hard to beat this because increasing $\chi_{\rm r}^{(3)}$ farther is generally accompanied by a reduction of the bandwidth and therefore the product $\chi^{(3)}{\cal E}^2$, see \eref{eq:photon}, which is a more useful figure of merit of the useful non-linearity, is not improved.\footnote{Very large $\chi^{(3)}$ coefficients are possible in liquid crystal devices (see e.g. S. Residori~\etal{} \cite{residori08}) but with a considerably lower bandwidth ($\sim$kHz) than atomic systems ($\sim$MHz).}

\subsection{Cooperative Phenomena}

To enhance the near resonant optical non-linearity we consider processes that are no longer mediated directly by the interaction between light and matter at the single particle level, but indirectly by interactions between light induced excitations. We describe such interactions as cooperative because they involve the cooperative response of many atoms.\footnote{In a cooperative process the interaction between dipoles modifies the response of each individual dipole. The classic example is superradiance \cite{dicke54}.} Such cooperative processes not only result in non-linearities that are larger than the Kerr effect but also can be long range meaning that two light beams can interact even without a spatial overlap \cite{sevi11b}.

For an atomic system, the simplest coupling capable of producing cooperative behavior is the long range interaction between {atomic} dipoles. If one atom is located at the origin, a second atom at position $R$ will experience an interaction with the dipole electric field of the first atom, $\mrb{E}(R)$, with energy
\begin{eqnarray}
V_{\rm dd}=-\frac{3\hbar\Gamma}{4}\left[\frac{1}{kR}\sin^2\theta+\left(\frac{1}{(kR)^3}-\frac{\ii}{(kR)^2}\right)(3\cos^2\theta-1)\right]\e^{-\ii kR}~,
\end{eqnarray}
where $\theta$ is the angle between the orientation of the atomic dipole and direction $R$. The two atoms can be prepared in pair states $\ket{gg},\ket{eg},\ket{ge}$ and $\ket{ee}$, however states $\ket{eg}$ and $\ket{ge}$ are degenerate and couple via the dipole--dipole interaction. The modified eigenstates of the interaction Hamiltonian are the Dicke \cite{dicke54} states $\ket{\pm}=(\ket{eg}\pm\ket{ge})/\sqrt{2}$. It can be shown that the effect of the coupling between the two atoms due to $V_{\rm dd}$ is to give rise to an energy level splitting (or shift) $E_{12}$ and broadening $\Gamma_{12}$ of the two states associated with the real and imaginary parts respectively \cite{lehmberg70,protsenko06}, i.e.,
\begin{subequations}\label{eq:D12}\begin{flalign}
\hbar\Gamma_{12} &= - 2\Im\{V_{\rm dd}\},  \\
\hbar\Delta_{12} &= 2\Re\{V_{\rm dd}\},
\end{flalign}\end{subequations}
where the modified energy and decay rates of the $\ket{\pm}$ states are given by
\begin{subequations}\label{eq:D12}\begin{flalign}
E_{\pm} &=\hbar\omega_0\pm E_{12},\\
\Gamma_{\pm} &=\Gamma\pm\Gamma_{12}.
\end{flalign}\end{subequations}

\begin{figure}[!t]
\begin{center}
\includegraphics[width=9cm,angle=0]{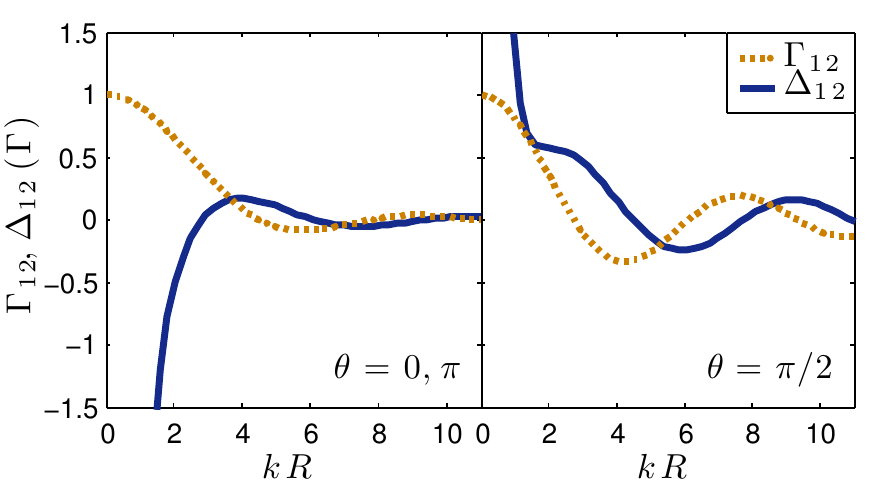}
\caption[]{Dipole--dipole induced broadening ($\Gamma_{12}$) and level shift ($\Delta_{12}$) for a pair of atoms as a function of separation for two relative orientations of their dipoles.}
\label{fig:dicke}
\end{center}
\end{figure}

The resulting splitting and decay rates are plotted in \fref{fig:dicke} for two different orientations as a function of $kR$. This shows that for $1\lesssim kR\lesssim10$ the dominant effect of the dipole--dipole coupling is to modify the spontaneous lifetime of states $\ket{\pm}$ to be faster or slower than $\Gamma$, known as superradiance and subradiance respectively. This has been demonstrated experimentally for a pair of trapped ions with variable separation \cite{devoe96}. For $kR<1$ the energy shift diverges, which can be exploited to realize a cooperative optical non-linearity that is enhanced by the number of atoms that cooperate in the interaction, i.e., $\chi^{(3)}\propto N^2$, resulting in a non-linear density dependence.

One of the challenges associated with observing this non-linearity for an optical transition is that it requires an interatomic separation $R\sim100$~nm, corresponding to relatively high densities $\sim10^{15}~$cm$^{-3}$. This has been achieved by probing a high temperature atomic vapor layer {of nanometer thickness} \cite{keaveney12} or using molecules in an organic crystal \cite{hettich02}, however an alternative approach is to use Rydberg states.

The advantage of Rydberg atoms is their extraordinarily large dipole moments for transitions between nearby Rydberg states. The dipole moment scales as the principal quantum number $n^2$. For $n > 40$ the inter-Rydberg transitions associated with large dipole moments have wavelengths of order millimeters. Consequently, the spatial extent of a cold atomic ensemble can be much less than the wavelength, i.e., $kR \ll1$ and the level shift dominates over superradiant effects. Whereas we required $N > 10^{15}$~cm$^{-3}$ for $kR > 1$ for dipole--dipole effects to dominate on an optical transition, this is reduced to $N > 10^9$~cm$^{-3}$ for Rydberg states. This hierarchy of length scales allows us to operate in a regime where dipole--dipole interactions between Rydberg levels are large but dipole--dipole interactions on the optical transition are small, i.e.,
\begin{equation}
\lambda_{\rm opt} < N^{-1/3}<\lambda_{\rm Ryd}~.
\end{equation}
Note that the dipole moment for optical transitions between the ground state and a Rydberg state is very small. Fortunately EIT allows us to combine the large level shifts associated with the strong dipole--dipole interactions between Rydberg states with the strong light coupling on optical transitions.

In this review we focus on the cooperative atom--light interaction arising in Rydberg systems {that can involve of the order of one thousand atoms}. The cooperative optical non-linearity occurring in a Rydberg ensemble was first reported by Pritchard {\it et al}. in 2010 \cite{prit10}. A $\chi_{\rm r}^{(3)}$ similar in magnitude to that reported in the work of Hau {\it et al} was observed, but at a much lower density. Scaling to the same density the non-linearity in the Rydberg system is 5 orders of magnitude larger than for a conventional EIT medium. In this case we have a non-linearity sufficiently large to enter the quantum regime where the classical optical Kerr like description breaks down, see \fref{fig:chi3}.

In the next section we consider how this large non-linearity arises in a Rydberg ensemble.

\section{Rydberg EIT}\label{sec:Rydberg}

\subsection{Dipole Blockade}\label{sec:blockade}
Rydberg states are highly excited states of a valence electron in an atom or molecule, resulting in hydrogenic scaling laws dependent upon the principal quantum number, $n$. The most useful property of the Rydberg states is the orbital radius $\propto n^2$ which is responsible for their enormous dipole moments, resulting in an exaggerated response to electric fields with a DC polarizability $\alpha\propto n^7$. As discussed above, combining this large dipole moment with the millimeter energy spacing between Rydberg levels such that $kR\ll1$ leads to strong dipole--dipole interactions between Rydberg atoms which are manifest as a position dependent level shift, $V(R)$. Typically experiments are performed in the Van der Waals regime where $V(R)=-C_6/R^6$, however it is possible to use an external electric or microwave field to realize a longer range resonant dipole--dipole coupling $V(R)=C_3/R^3$.\footnote{For a detailed discussion of Rydberg atom dipole--dipole interactions see Saffman \etal{} \cite{saffman10}. A table of Van der Waals $C_6$ coefficients for the alkali atoms can be found in Singer \etal{} \cite{singer05}.} This level shift can be either positive or negative and may depend on angle, resulting in controllable attractive or repulsive interactions. For example, for the specific case of Rb coupling to the zero angular momentum Rydberg series ($n$s states), one obtains an isotropic positive shift that is independent of angle. In this case the van der Waals coefficient depends on the eleventh power of the principle quantum number $C_6\propto n^{11}$. This high order dependence on $n$ enables convenient tuning of the interaction strength.

\begin{figure}[!t]
\begin{center}
\includegraphics[width=9cm,angle=0]{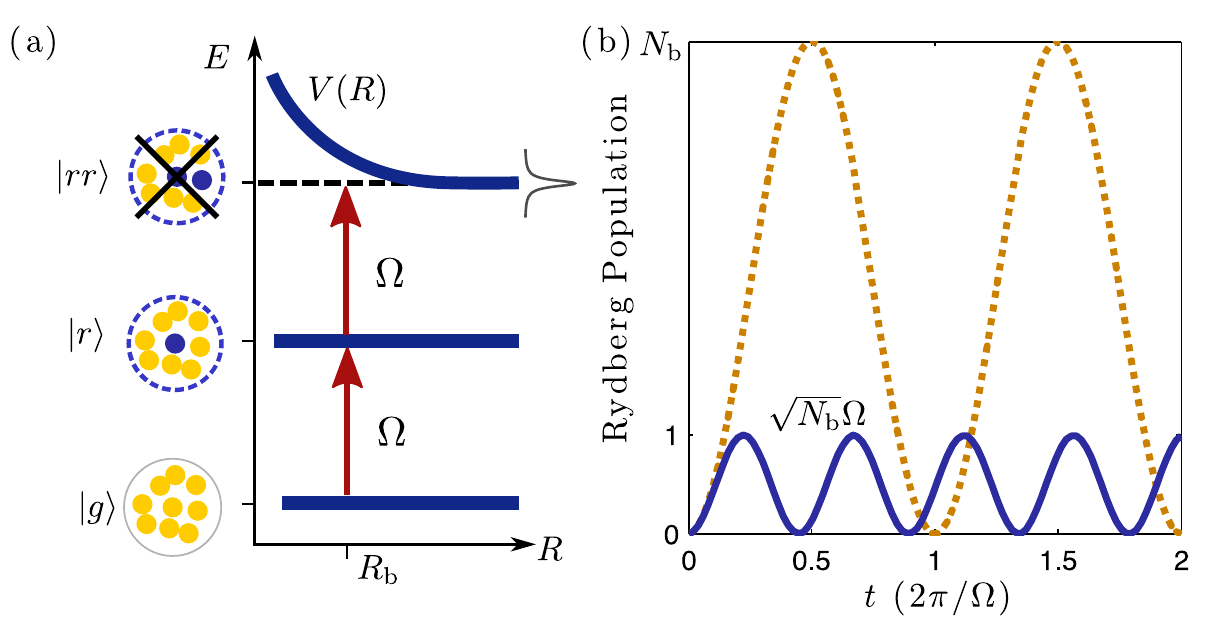}
\caption[]{Schematic diagram showing Rydberg blockade for two atoms. (a) In an ensemble of $N$ within a volume known as the blockade volume only one Rydberg excitation is allowed due to the dipole--dipole induced shift of the doubly excited state $\vert rr\rangle$. (b) The blockade process leads to a speed up of the Rabi oscillation by a factor $\sqrt{N_{\rm b}}$.}
\label{fig:blockade}
\end{center}
\end{figure}

An important consequence of this position dependent level shift is dipole--blockade, which prevents multiple Rydberg excitations within a sphere of radius $R_{\rm b}=(C_6/\hbar\Omega)^{1/6}$, where $\Omega$ is the Rabi frequency of the resonant coupling from the ground state $\ket{g}$ to Rydberg state $\ket{r}$ as illustrated in \fref{fig:blockade}(a). Typically the blockade radius is of order microns, for example for the Rb $60S_{1/2}$ state with $\Omega/2\pi=1$~MHz, $R_{\rm b}\sim7~\mu$m. Blockade therefore enables deterministic creation of the singly excited collective state
\begin{equation}
\ket{\Psi} = \frac{1}{\sqrt{N_{\rm b}}}\displaystyle\sum_{i=1}^{N_{\rm b}}\ket{g_1g_2g_3\ldots r_i\ldots {g_{N}}_{\rm \!b}}~,
\end{equation}
where $N_{\rm b}=N(4/3)\pi R_{\rm b}^3$ is the number of atoms per blockade sphere. Each blockade sphere now acts as an effective two-level atom or ``superatom" \cite{vuletic06}, with the Rabi frequency between state $\ket{g_1\ldots g_{N{\rm b}}}$ collectively enhanced by a factor $\sqrt{N_{\rm b}}$ as shown in \fref{fig:blockade}(b).

Interest in Rydberg atom dipole blockade was piqued at the start of 2000 by the proposal of Jaksch \etal{} \cite{jaksch00}, closely followed by Lukin \etal{} \cite{lukin01}, who proposed exploiting this deterministic process to realize fast atomic quantum gates for neutral atoms. Subsequently, significant theoretical and experimental studies have extended this work, with the first demonstrations of a two-atom C--NOT gate \cite{isenhower10} and entanglement generation \cite{wilk10} in 2010. A detailed review of quantum information processing with Rydberg atoms can be found by Saffman \etal{} \cite{saffman10}. In this work however, we focus on exploiting the effects of dipole blockade on the light field.

\subsection{Rydberg EIT}\label{sec:RydEIT}

For applications in non-linear optics one can combine the extraordinary properties of Rydberg states, including dipole blockade, with a strong atom--light coupling using the technique of EIT, discussed in \sref{sec:EIT}.

\begin{figure}[!t]
\begin{center}
\includegraphics[width=3cm,angle=0]{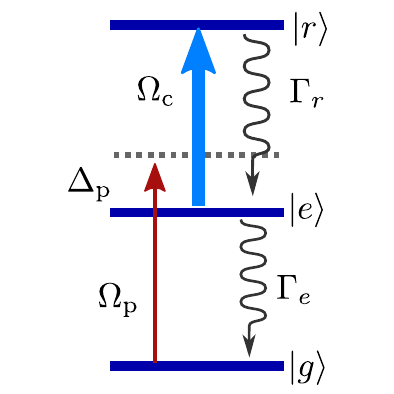}
\caption[]{The level scheme used in Rydberg electromagnetically induced transparency (EIT). }
\label{fig:3levels}
\end{center}
\end{figure}

In Rydberg EIT a three-level ladder system is used where a weak probe laser drives the strong optical transition from the ground state $\ket{g}$ to excited state $\ket{e}$ with Rabi frequency $\Omp$ whilst a classical coupling field with Rabi frequency $\Omc$ couples $\ket{e}$ to the Rydberg state $\ket{r}$, as shown in \fref{fig:3levels}. For the non-interacting case, on the EIT resonance the atoms are prepared in the dark state
\begin{eqnarray}
\vert\psi\rangle_{\rm dark}&= &\cos\theta\vert g\rangle-\sin\theta\vert r\rangle~,\end{eqnarray}
where $\tan\theta=\Omega_{\rm p}/\Omega_{\rm c}$. This dark state is not coupled to the probe laser, resulting in the transparency window shown in \fref{fig:chi}(b). EIT can therefore be used as an optical probe of the Rydberg energy level, which combined with the large DC Stark shift is useful for applications in electrometry as will be discussed in \sref{sec:Efield}.

\begin{figure}[!t]
\begin{center}
\includegraphics[width=9cm,angle=0]{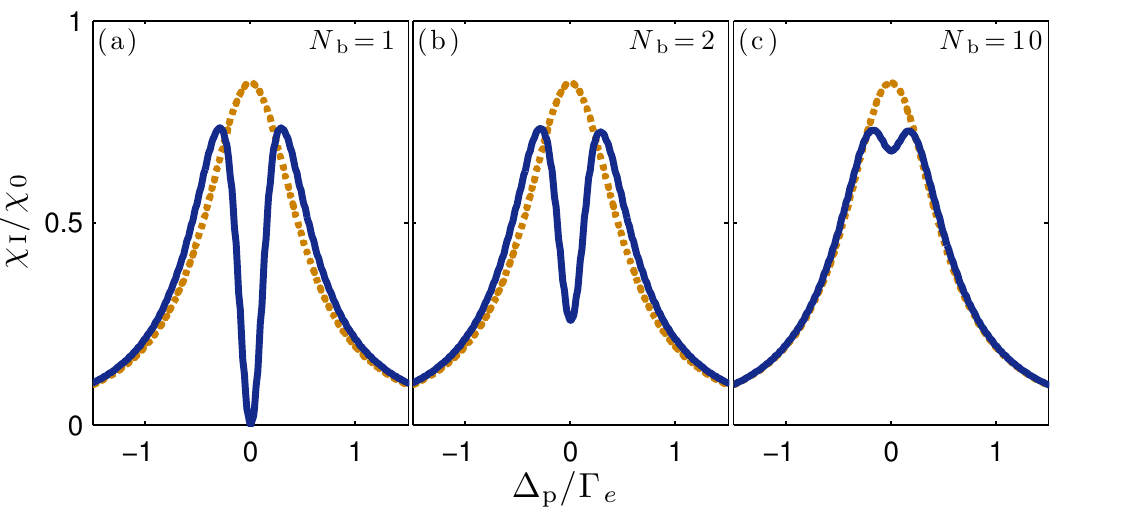}
\caption[]{The change in the optical response per atom in a blockaded Rydberg EIT medium (solid) as the number of atoms per blockade sphere is increased, compared to the susceptibility of a two-level atom (dashed). Calculated for $\Omp=\Ge/4, \Omc=\Ge/2$.}
\label{fig:interacting}
\end{center}
\end{figure}

\begin{figure}[!h]
\begin{center}
\includegraphics[width=9cm,angle=0]{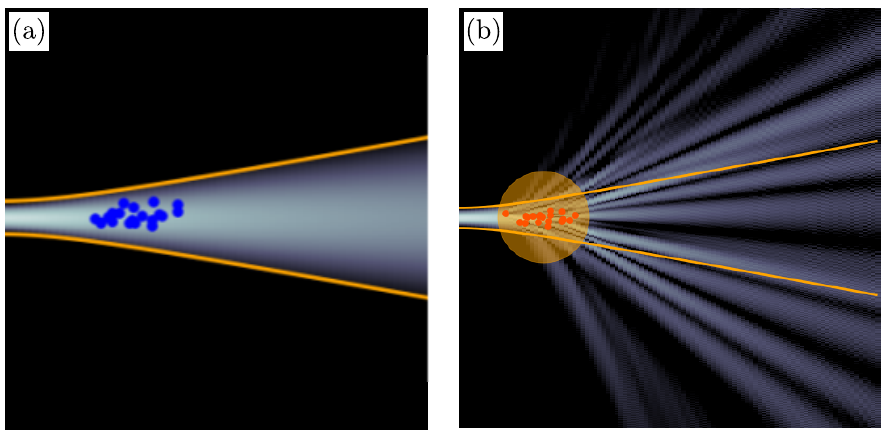}
\caption[]{Light propagation through a single blockade sphere. A single Rydberg excitation switches the neighboring atoms out of the transparent EIT dark state (a) into resonant two-level scatterers (b), suppression transmission in the mode of the probe laser. \label{fig:scatter}}
\end{center}
\end{figure}

As $\Omega_{\rm p}$ increases, more population is transferred to the Rydberg state and the Rydberg atoms begin to interact via induced dipole--dipole interactions, where the blockade radius is now defined in terms of the EIT linewidth $\Delta_{\rm EIT}$ as $R_{\rm b}=\sqrt[6]{C_6/\hbar\Delta_{\rm EIT}}$. The effect of dipole blockade on EIT is that a single Rydberg excitation can switch the optical response of the surrounding atoms within each blockade sphere from the transparent EIT condition to the resonant two-level scattering limit. This can be seen from considering the blockaded dark state for a pair of atoms given by \cite{moller08}
\begin{eqnarray}\label{eq:drk}
\vert\psi\rangle_{\rm dark}&= &\cos^2\theta\vert gg\rangle-\sin\theta\cos\theta(\vert gr\rangle+\vert gg\rangle)+\sin^2\theta\vert ee\rangle~,\end{eqnarray}
{where the doubly excited Rydberg state $\ket{rr}$ is replaced by $\ket{ee}$, which has the effect of increasing the scattering on resonance.} This is illustrated in \fref{fig:interacting} where we plot the imaginary part of the susceptibility per atom as we add more atoms per blockade sphere. This reveals a clear suppression of the EIT signature that tends towards the resonant absorption of a two-level atom as $N_{\rm b}$ increases. In standard optical media, the optical response of a single atom remains unchanged as a function of density. For the blockade effect however, adding more atoms leads to a modification of the single-atom response, a clear signature of a cooperative effect. In \fref{fig:scatter} we illustrate the light propagation when one Rydberg excitation turns the ensemble into two-level scatterers, demonstrating the removal of light from the probe beam mode.

This controlled switching between three-level and two-level optical responses as the Rydberg fraction increases is the mechanism behind the optical non-linearity. As $N_{\rm b}$ atoms are involved, on resonance the maximum non-linearity is enhanced by a factor of $N_{\rm b}$, i.e.
\begin{eqnarray}
\chi^{(3)}&\sim &N_{\rm b}\chi^{(1)}\frac{d_{\rm ge}^2}{(\hbar\Omega_{\rm c})^2}~,\end{eqnarray}
resulting in a quadratic density scaling. A more detailed analysis by S. Sevin\c{c}li \etal \cite{sevi11b} reveals that exactly on resonance  $\chi^{(3)}$ has equal real and imaginary parts. Off resonance the real part can be larger than the imaginary part allowing the possibility of low loss photon--photon interactions.

For typical parameters such as a Rydberg state with principle quantum number $n=60$, $R_{\rm b}=5~\mu$m and density $N=3\times 10^{12}~{\rm cm}^{-3}$, $N_{\rm b}=1500$ and one obtains a non--linearity $\chi_{\rm r}^{(3)}\sim 5\times 10^{-2}~{\rm V}^{-2}{\rm m}^2$, see \fref{fig:chi3}. As the photon field can be of order {${\cal E}\sim 30$~Vm$^{-1}$} the non--linearity can be saturated at the single photon level and the classical description breaks down, which will be considered in \sref{sec:qo}. In the following section, we review the experimental progress in the field of Rydberg EIT in the regime of classical Kerr non-linearities.

\section{Electromagnetically induced transparency in cold Rydberg gases}\label{sec:expt}

Using cold atomic gases to study Rydberg physics confers several benefits over using thermal vapors or atomic beams. For example one can achieve long coherence times in cold trapped atoms and it is possible to prepare all atoms in a particular quantum state. Also, because the average distance moved by an atom during an experiment is typically much less than the interparticle separation, the motion of the atoms can be neglected. Mourachko {\it et al.} coined the term ``frozen Rydberg gas" to describe this regime and observed many-body effects in cold atoms excited to Rydberg states \cite{mour98}. Since then frozen Rydberg gases have allowed the study of many different phenomena, for example the resonant dipole energy transfer between Rydberg states  \cite{ande98} and the processes by which cold Rydberg gases spontaneously form ultracold plasmas  \cite{kill99,pohl03}.  Other exotic forms of matter such as long range Rydberg molecules  \cite{bend09} are formed when the interaction between a Rydberg atom and a ground state atom leads to a bound state. Cold gases also allow for precision measurement, for example the quantum defects of Rydberg series in Rb were measured to unprecedented precision using millimeter wave spectroscopy of laser cooled atoms \cite{li03}.

As discussed in \sref{sec:blockade}, one of the most interesting aspects of Rydberg physics is the dipole blockade. In laser cooled gases where the typical density is around $10^{10}$~cm$^{-3}$, the inter--atomic separation can be less than the blockade radius and several experiments observed evidence of excitation blockade  \cite{singer04,tong04,cubel05}. At higher densities, a conclusive demonstration was shown by Heidemann {\it et al.} who observed the $\sqrt{N_{\rm b}}$ collective enhancement of the excitation of Rydberg states in a dense ultra-cold cloud  \cite{heid07}. Laser cooling also allows experiments on isolated atoms. Two independent groups observed blockade between two individual atoms trapped in optical dipole traps  \cite{urba09,gaet09} separated by a few microns, followed by subsequent demonstrations of deterministic entanglement \cite{wilk10,isenhower10}. These experiments can be considered the first step on the path towards quantum information with Rydberg atoms \cite{saffman10}.

The phenomenon of dipole blockade is theoretically not exclusive to cold gases, however for thermal vapors where it is possible to achieve much higher densities the blockade radius is reduced by a factor of $1/\sqrt[6]{\nu_{\rm Doppler}}$, where $\nu_{\rm Doppler}$ is the frequency width due to Doppler broadening. This, combined with a requirement to probe the cloud on nanosecond timescales to achieve the equivalent ``frozen" state, has meant blockade in vapors cells has so far eluded detection, although coherent Rabi oscillations have been observed recently \cite{huber11}.

In this section we discuss how cold atomic gases can be used to observe non-linear optical phenomena involving Rydberg states and how the blockade affects these processes. We present experiment and theory relating to classical non-linear optical phenomena.

\subsection{Coherent Optical Detection of Rydberg States}\label{EITexpt}

\begin{figure}[t!]
\centering
	\includegraphics[width=10cm]{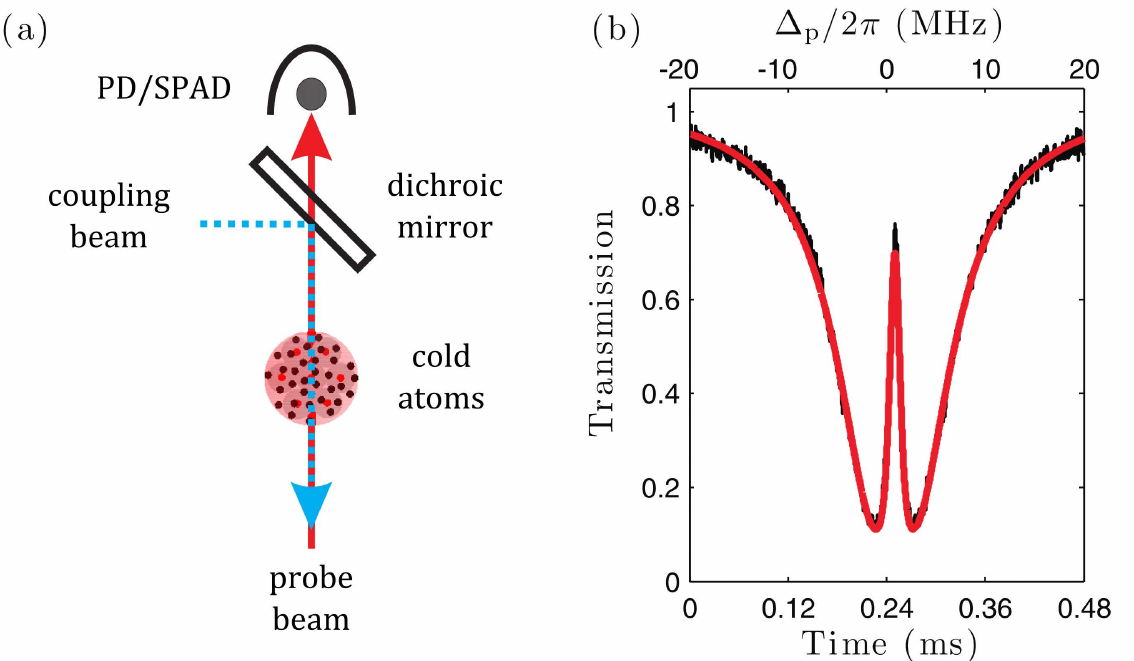}
	\caption{\label{fig:fig1} (a) Schematic of experimental setup. Spectroscopy is performed on a cold atomic sample by scanning the probe beam through resonance. (b) Spectrum for the 46$D_{5/2}$ state in $^{85}$Rb.}
\end{figure}

Electromagnetically induced transparency with Rydberg states using the ladder configuration shown in \fref{fig:3levels} was first performed in room temperature Rb vapors by Mohapatra {\it et al.}  \cite{mohapatra07}. This demonstration was the first coherent optical detection of Rydberg states and offered a non-destructive alternative to ion detection using micro-channel plates or channeltrons.

Shortly afterwards Weatherill {\it et al.}  \cite{weat08} demonstrated Rydberg EIT on cold atoms. The experiments were performed on a cold Rb cloud released from a magneto-optical trap (MOT) with a typical temperature of 20~$\mu$K. Figure~\ref{fig:fig1}(a) shows a schematic of the experimental setup. Counter-propagating coupling and probe beams which cross at a dichroic mirror, pass through the sample of atoms. The transmission of the probe laser is measured at a photodiode (PD) or single photon avalanche detector (SPAD). By quickly scanning the probe laser through resonance an EIT spectrum is obtained, Fig~\ref{fig:fig1}(b) shows an EIT spectrum for the 46$D_{5/2}$ state. In the weak probe regime ($\Omega_{\rm p}\ll \Omc$) repeated scanning across the resonance whilst the coupling laser is continuously on shows no significant degradation in signal. This technique therefore provides a direct and non-destructive probe of Rydberg states in cold atoms. The observed transparency features are narrow ($<$ 1~MHz), with the width of the resonance limited by the residual two-photon laser linewidth of approximately 200~kHz.

This coherent probe of Rydberg states opens the door for new studies, for example, following similar methods Piotrowicz \etal{} performed systematic measurements of dipole moments of transitions to Rydberg levels in cold Rb  \cite{piot11}. By measuring the Autler-Townes splitting as a function of coupling laser power, dipole moments for transitions to a range of Rydberg states were determined and the results were compared to a variety of theoretical models.

\subsection{Sensitivity to electric fields}\label{sec:Efield}
\subsubsection{Giant Electro-Optic Effect}
\begin{figure}[t!]
\centering
	\includegraphics[width=11cm]{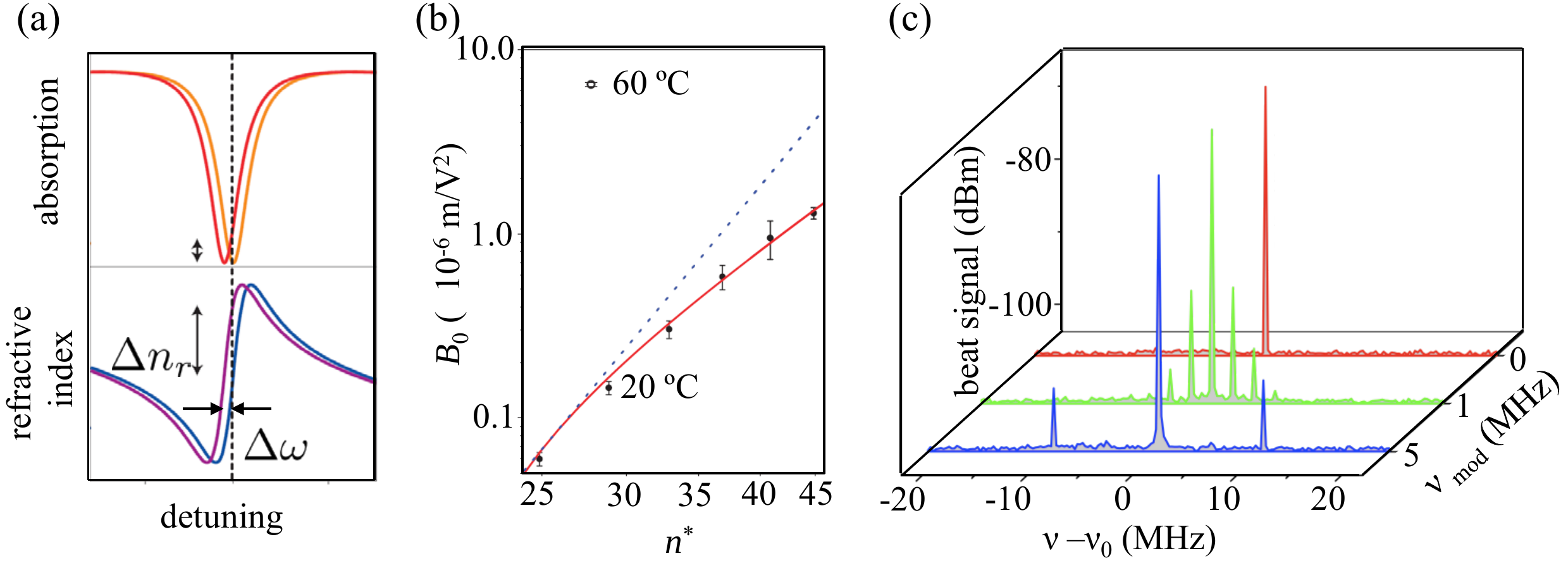}
	\caption{\label{fig:EOM} Giant electro-optics effect (a) Change in refractive index $\Delta n$ is approximately linear with detuning and the frequency shift is proportional to the square of electric field (b) Measured Kerr constant in a vapor cell at 20 $^{\circ}$C (filled circles) and 60 $^{\circ}$C (open circle) as a function of reduced principal quantum number, $n^{*}$ . A theoretical fit (solid line) is shown for constant laser power  (c) Power spectrum of the signal light transmitted through the dark-state ensemble. The sidebands are generated by modulating an applied electric fields of amplitude 2.4 V/cm and frequency $\nu_{\rm mod}$.}
\end{figure}

As described in \sref{sec:RydEIT}, by using EIT we can combine the strong atom-light coupling of optical transitions with the large Stark shifts associated with highly polarizable Rydberg states. In this configuration, an externally applied electric field $E$ causes a DC or AC Stark shift of the Rydberg state energy $\Delta\omega=-(1/2)\alpha E^2$ which shifts the dispersion profile of the EIT resonance leading to a change of $\Delta n_{\rm r}$ in the refractive index, as illustrated in \fref{fig:EOM}(a). This results in an extremely large DC or low frequency Kerr effect of the form
\begin{equation}
\Delta n_{\rm r}= B_0 \lambda E^2~,
\end{equation}
where $B_0$ is the electro-optic Kerr coefficient which is equivalent to $\chi^{(3)}/\lambda$.

The first demonstration of this `giant electro-optic' effect was reported by Mohapatra {\it et al.} \cite{mohapatra08} in thermal Rb vapors. Kerr coefficients many orders of magnitude larger than typical Kerr media were demonstrated, as shown in \fref{fig:EOM}(b). The coefficients, $B_0$ were determined using an interferometer to measure the phase shifts, $\Delta \phi$ induced by an applied electric field, $E$ since from \eref{eq:phi} $\Delta \phi = 2 \pi B_0 \ell |E|^2$, where $\ell$ is the length of the medium, in this case a 7.5~cm vapor cell. A further demonstration of this effect was illustrated by producing sidebands on the probe field at frequency $\nu_{\rm mod}$ of the modulated electric field. \Fref{fig:EOM}(c) shows the power spectrum of light transmitted through a dark state ensemble for a range of different modulation frequencies. In addition, it was shown that the bandwidth of the low frequency Kerr response was determined by the EIT transient response. It is worth noting that much larger Kerr coefficients would be accessible using cold gases due to the higher group index.

\subsubsection{Electrometry}
The enhanced electric field sensitivity of Rydberg states makes them ideal candidates for precision electrometry, with sensitivities of 20~$\mu$V/cm achieved using Rydberg states of Krypton in an atomic beam by Osterwaler \etal{} \cite{osterwalder99}. Their method required use of selective field ionization, however as demonstrated above EIT enables a non-destructive probe of the Rydberg energy level, making it possible to study electric field dynamics in real-time using a single atomic sample.

\begin{figure}[t!]
\centering
	\includegraphics[width=8cm]{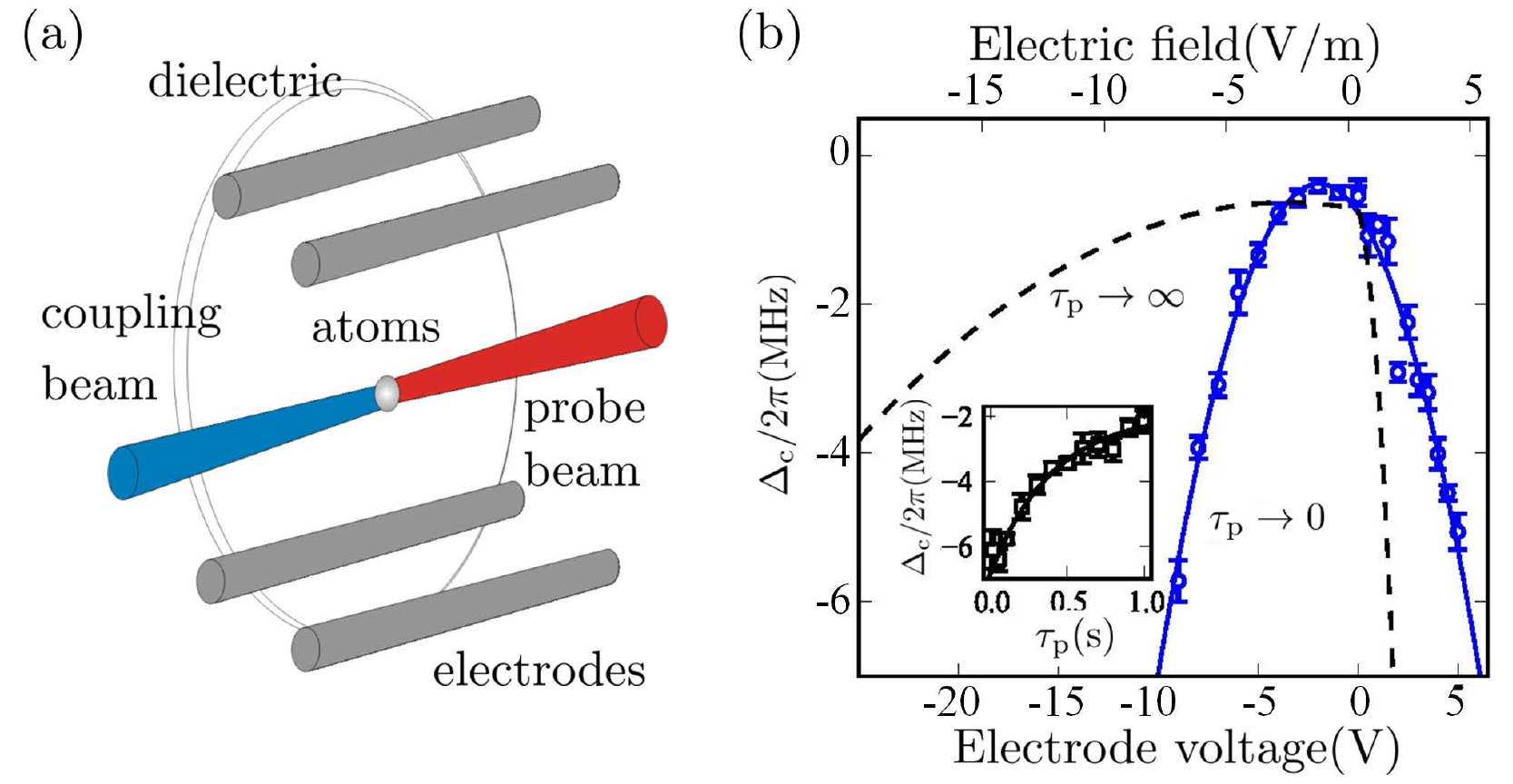}
	\caption{\label{fig:abel} (a) Experimental setup for measuring electric fields close to a dielectric surface using Rydberg EIT. (b) The electric field dependence of the EIT resonance when the electrode voltage pulse is short, $\tau_{\rm p} \rightarrow 0$ (solid line) and the pulse is long $\tau_{\rm p} \rightarrow \infty$. (dashed line). The inset shows the level shift as a function of voltage pulse length with the voltage fixed at 9~V.}
\end{figure}

An interesting application of this technique is to probe the stray fields close to a surface, where the electric field can be determined from measuring the DC Stark-shift of the EIT resonance, as first demonstrated by Tauschinsky \etal{}  \cite{taus10}. Using the same idea Abel {\it et al.} used EIT to measure the electric field close to the surface of a dielectric \cite{abel11}. \Fref{fig:abel}(a) shows the experimental setup, with laser cooled Rb atoms probed at the center of an array of four electrodes located 12.5~mm away from the dielectric surface (an AR coated glass window). The Rydberg energy level shift was measured as a function of the voltage applied to the electrodes and is shown in \fref{fig:abel}(b). The measured Stark map shows drastically different behavior depending on how long the electric field is applied, suggesting that the properties of the dielectric play a role in the local electric field. The surface dipoles of adsorbate atoms create a dipolar field that is proportional to the applied field and always points away from the surface. The behavior of the shift in the EIT resonance can be modeled by,
\begin{equation}
\frac{\Delta_{\rm c}}{2\pi}=-\frac{1}{2}\alpha [E + \beta(\tau_{\rm p})|E|+E_0]^2,
\label{eq:elec_shift}
\end{equation}
where $E$ is a constant background field and $\beta(\tau_{\rm p})|E|$ is an additional field which depends upon the electrode pulse duration, $\tau_{\rm p}$.

The data points (solid line) show the electric field dependence of the resonance when the electrode voltage pulse is short, $\tau_{\rm p} \rightarrow 0$ and the dashed line is a functional fit of \eref{eq:elec_shift} to the data for $\tau_{\rm p} \rightarrow \infty$. For pulsed electric fields the Rydberg energy level shift exhibits a quadratic dependence on the applied voltage and is fit using \eref{eq:elec_shift} with $\beta=0$. The inset to \fref{fig:abel}(c) shows the behavior of the level shift as a function of voltage pulse time for a fixed voltage of 9~V. The shift shows an exponential decay with time constant $\sim$ 1~s which is a measure of the time response of the adsorbates in the dielectric.

In the work of Tauschinsky \etal{} EIT was used to measure the local electric fields arising from Rb atoms adsorbed on the surface of an atom chip  \cite{taus10}, where transmission is recorded using a CCD camera to obtain a spatially resolved electric field sensitivity of 0.1 V/cm with a 7~$\mu$m resolution. This technique allows the field to be mapped out at all points around the surface simultaneously, making it sensitive to electric field gradients. Gaining a good understanding of the properties of Rydberg atoms close to surfaces is of considerable interest to quantum information schemes proposing to combine Rydberg atoms with scalable atom chip experiments  \cite{mull11,leun11}.

\subsection{Interaction effects}\label{interactions}
\subsubsection{Superradiant Cascade}
We have so far considered only the effect of externally applied fields on Rydberg EIT, but at sufficiently high atomic density and principal quantum number the dipole--dipole interactions between Rydberg atoms begin to play an important role and lead to a modification of the optical transmission. The range of atomic densities accessible in a typical cold atom experiment allow the exploration of these effects.

\begin{figure}[t!] 	
\centering
    \includegraphics[width=7cm]{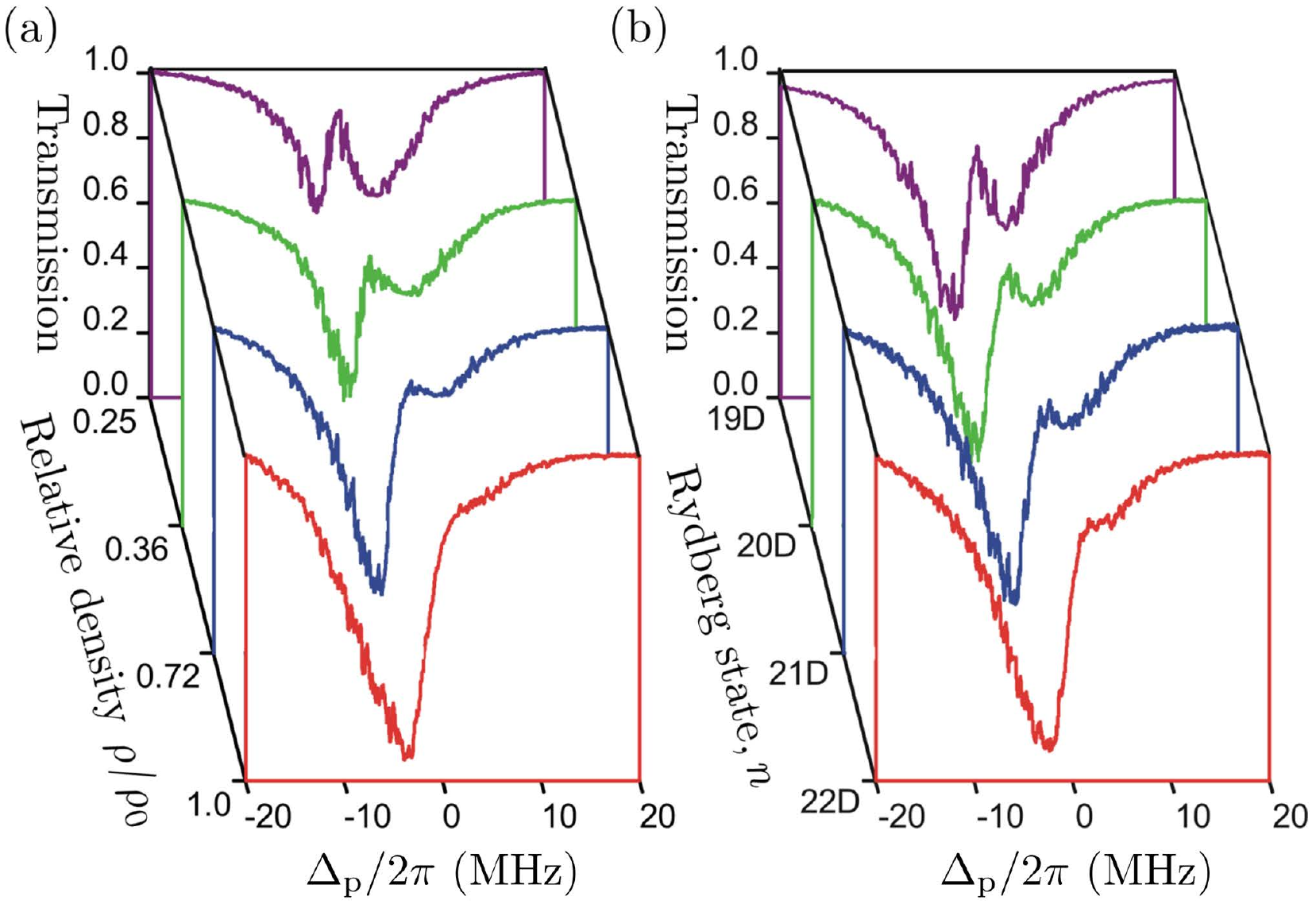}
	\caption{\label{fig:superradiance} (a) Transmission of a weak probe beam as a function of
$\Delta_{\rm p}$ for the $^{87}$Rb $ 5s \rightarrow 5P \rightarrow 26d$ system
for varying atomic density, $\rho$. (b) EIT spectra for $n$ = 19 - 22. The Rabi frequency of the
Rydberg transition is kept the same by varying the coupling laser power. The onset of density-dependent behavior is evident in the emergence of an EIT lineshape at lower values of $n$ and $\rho$.}
\end{figure}

As discussed in the introduction, dipole--dipole interactions between a pair of dipoles results in a modification to both the pair state energy level and the decay rate for $kR\le1$. For low-lying Rydberg states ($n\sim20$), the transition wavelength of the closest lying Rydberg states is approximately 1~mm, comparable to the size of a typical cold atom cloud trapped in a magneto-optical trap. As a result, there is a geometric enhancement of the cooperative coupling of the dipoles  \cite{rehler71} that results in the superradiant decay dominating over any effect of the level shift. As a consequence there is a rapid population transfer into close-lying Rydberg states, which may then in turn undergo superradiant decay. This effect is known as superradiant cascade  \cite{gros76}, which can be detected from a redistribution of population into a range of lower energy Rydberg states. Recent studies of Rydberg excitation in cold atoms have shown this to play an important role in the dynamics  \cite{wang07,day08}. In the initial experiments performing Rydberg EIT on cold atoms, this effect was observed through an abrupt density dependent loss as the probe laser is scanned across the two-photon resonance  \cite{weat08}, as illustrated in \fref{fig:superradiance} where the $kR$ dependence can be seen through the variation of the spectra with both (a) density (changing $R$) and (b) principal quantum number (changing $k$). In order to observe the more useful cooperative effects arising from dipole blockade, it is therefore necessary to use highly excited Rydberg states such that the interactions $\propto n^{11}$ dominate over the superradiant decay $\propto n^{-5}$.\footnote{This scaling for superradiant decay is obtained from $\Gamma\propto\omega^3d^2$, where the transition energy to the close lying $n-1$ states $\omega\propto n^{-3}$ and dipole moment $d\propto n^2$.}

\subsubsection{Cooperative Optical Non-linearity}

The dipole--dipole level shifts become relevant for Rydberg states in Rb for $n\approx40$, where the blockade radius $R_{\rm b}$ becomes comparable to the average nearest-neighbor separation between atoms in a laser cooled sample with a typical density of $10^{10}~$cm$^{-3}$, making studies of cooperativity in the blockade regime possible. The majority of work in this vain has been performed using S states in Rb because their repulsive interactions make them stable against ionizing collisions  \cite{amth07}, and there is also little coupling to adjacent pair states  \cite{pohl09,youn09}.

\begin{figure}[t!] 	
\centering
    \includegraphics[width=9cm]{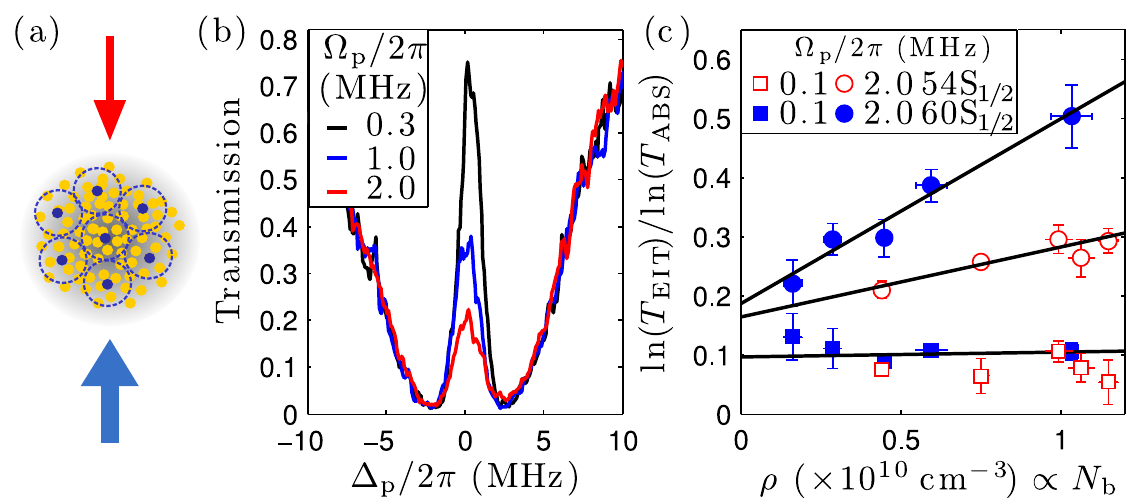}
	\caption{\label{fig:fig3} Cooperative optical non-linearity: (a) EIT is performed on an atomic ensemble composed of several blockade spheres (b) Transmission spectra for $60S_{1/2}$ revealing strong suppression of transparency due to Rydberg blockade. (c) Dependency of the optical depth with atomic density for weak (squares) and strong (circles) probe Rabi frequency. The values are divided by the probe-only values to remove trivial linear density scaling. States 60S and 54S are compared.}
\end{figure}

The effect of dipole blockade in Rydberg EIT is to create a cooperative optical non-linearity where the presence of a single Rydberg excitation causes the surrounding atoms to switch to resonant two-level scattering, as discussed in \sref{sec:RydEIT}. This effect was investigated by Pritchard \etal{} \cite{prit10} by performing EIT on an atomic ensemble composed of many blockade spheres, as illustrated in \fref{fig:fig3}(a). Transmission spectra obtained with a coupling to the 60$S_{1/2}$ state are shown in \fref{fig:fig3}(b), with $R_{\rm b}=6~\mu$m corresponding to an average of $N_{\rm b}\sim11$ atoms per blockade sphere.  These data clearly demonstrate the anticipated suppression of the resonant transparency as the probe Rabi frequency is increased (and hence Rydberg state population), as predicted in \fref{fig:interacting}. An important feature of the observed lineshape is that the suppression is not accompanied by a frequency shift or broadening, ruling out ionization or dipole--dipole dephasing as the mechanism responsible for the non-linearity.

Conclusive evidence of a cooperative effect requires observation of a non-linear density dependence in the susceptibility caused by the optical response of each atom being dependent on the presence of its neighbors. This evidence is provided by data in \fref{fig:fig3}(c) which shows the EIT optical depth at $\Delta_{\rm p}=0$ scaled by the probe-only optical depth to give the optical response per atom. For the weak probe data (squares), the optical response is density independent as expected for a standard optical non-linearity. However, at stronger probe powers (circles) there is a clear linear dependence on the number of atoms per blockade sphere, $N_{\rm b}$, which is modified by varying the ground state atom density or the size of the blockade radius through choice of Rydberg state. The gradient of the density scaling for the 60S and 54S states have a ratio of 2.6$\pm$0.7 which is consistent with the expected value of 2.0 obtained from the scaling of $R_{\rm b}\propto n^{6.25}$. From the linewidth of the EIT features, an upper bound of 110~kHz was deduced for the dephasing rate of the blockade spheres. This linewidth is not limited by atomic properties but rather the linewidth of the lasers. In the experiments described here, the lasers are stabilized using Rydberg EIT in a vapor cell  \cite{abel09} and results in a linewidth of 110~kHz. Further reduction of the laser linewidth, perhaps by stabilizing to a cavity, would yield lower dephasing rates. This limit on the dephasing between blockade spheres is lower than the interaction-induced dephasing rates measured by Raitzsch {\it et al.} from spin echo and EIT techniques on cold Rydberg atoms  \cite{rait09}, and demonstrates that the blockaded ensemble is suitable for applications in quantum information where the strong interactions permit gates on microsecond timescale \cite{jaksch00}.

For low densities where $N_{\rm b}\sim3$, excellent theoretical agreement is achieved with the observed suppression of the EIT lineshape using a complete model of the interacting 3-atom system, however extension to many  atoms per blockade sphere becomes intractable. Previous models of the many-body interactions relied on treating the interactions as a mean-field shift on the Rydberg state of each atom, obtained by a pair-wise summation over $V(R_{ij})$ for all the surrounding atoms \cite{tong04,weimer08,chotia08}. For the repulsive interactions of the $S$ states, this predicts a frequency shift and broadening of the EIT resonance for increasing probe power which is not observed experimentally. Instead, due to the cooperative nature of the interaction it is necessary to include the correlations between all atoms in the system. This point was first demonstrated in the work of Schempp \etal{} \cite{sche10}, who performed complementary experiments studying the effect of dipole--dipole interactions on narrow coherent population trapping (CPT) using the same three-level system and required inclusion of two- and three-body correlations to reproduce their measured frequency dependent Rydberg atom number. Significant theoretical progress has been made by Ates \etal{}, who developed a Monte Carlo method \cite{ates11} that accounts for all orders of atomic correlations, facilitating modeling of experimental data at high densities. This approach has been shown to give good agreement with both the EIT data presented here and the CPT data of Schempp \etal{}, predicting a universal scaling law between the Rydberg excitation fraction and the resonant transmission  \cite{sevi11}. Subsequently, this has been extended by Sevin{\c c}li {\it et al.} \cite{sevi11b} to derive analytic expressions for the non-linear susceptibility due to dipole--dipole interactions, revealing that Rydberg gases are an ideal non-linear medium to study non-local wave phenomena. For example, for attractive interactions, modulation instabilities can occur at reasonable experimental parameters and give rise to stable bright solitons.

{Petrosyan \etal{} \cite{petrosyan11} present an alternative approach to modeling the optical propagation through the strongly interacting EIT medium by including an explicit coupling to the second-order correlation function of the probe field. Their model achieves excellent agreement with the data presented in \fref{fig:fig3}(a) and they also predict highly non-classical photon statistics associated with the resonant suppression.} Further theoretical analysis of Rydberg EIT involving photon--photon correlation effects will be covered in \sref{sec:qo}.

\subsection{Attractive Interactions}

To investigate the effect of the sign of the atom-atom interactions on optical transmission, studies were also performed using Rydberg states with attractive interactions (D states in Rb)  \cite{prit11}. In this case motional effects and ionization can play a role in the optical transmission making the optical non-linearity a function of time as well as optical field strength. This is due to the mechanical force arising from the attractive potential, which accelerates Rydberg atom pairs towards each other leading to ionizing collisions on timescales of order 10~$\mu$s \cite{gall08}.

\begin{figure}[t!]
\centering
	\includegraphics[width=9cm]{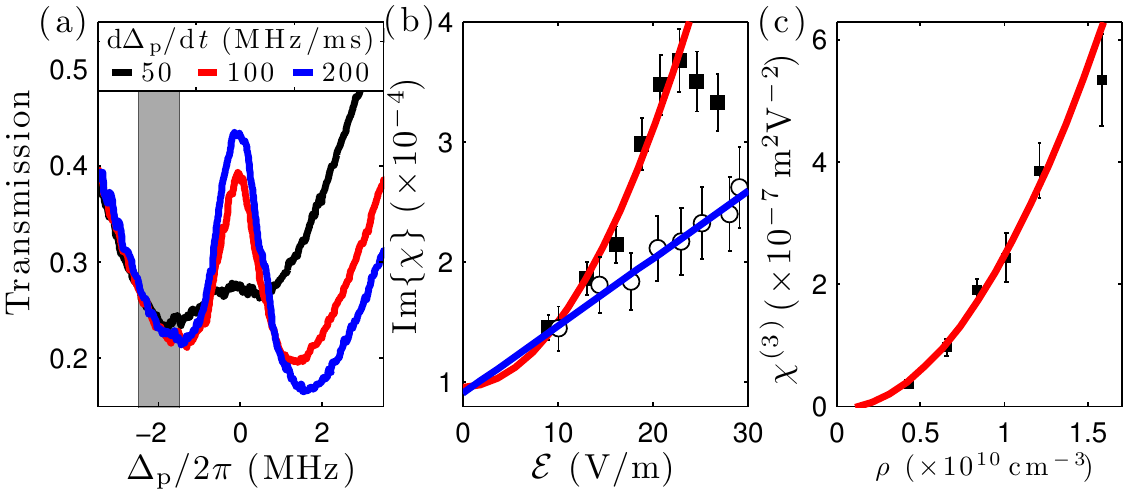}	
	\caption{\label{fig:fig4} {Rydberg EIT with attractive interactions. (a) EIT spectra for 58d as a function of scan rate reveals ionisation of the anti-blockade states excited in the gray shaded region dominate over cooperative interactions for slow scan speeds. (b) Resonant susceptibility as a function of probe electric field for positive (squares) and negative (circles) scan directions display a $\chi^{(3)}$ and $\chi^{(2)}$ dependence respectively. (c) Quadratic density dependence of $\chi^{(3)}$ extracted from fitting to the resonant susceptibility.}}
\end{figure}

Figure \ref{fig:fig4}(a) shows how the width and height of the transparency feature for the $5s \rightarrow 5p \rightarrow 58d$ system changes with the rate at which the probe laser frequency is scanned across the EIT resonance. For the slowest scan rate of 50~MHz/ms there is clear evidence of broadening, asymmetry and loss consistent with significant ionization of off-resonant pair states (which are located in the grey shaded region). These states are known as anti-blockade states, as they facilitate deterministic excitation from $\ket{gg}$ to $\ket{rr}$ via a two-photon transition at $\Delta_{\rm p}=V(R)/2$ without any population of the singly excited states. At faster scan speeds, the transparency remains approximately constant as ionizing collisions are far less likely to occur over the timescale of the experiment.

The direction of the probe frequency scan also plays a role at high $\Omega_{\rm p}$. For example, for positive scan direction there is strong suppression of the resonance whereas for a subsequent negative scan back through resonance, the EIT is revived. This trend reveals that the anti-blockade states could play an important role in determining the magnitude of the EIT suppression. A measurement of the imaginary component of susceptibility can be made from the transmission spectra using the relation $\chi_{\rm i}= -\log(T)/k \ell$, where $T$ is the EIT transmission on resonance, $\ell$ is the length of the atom cloud and $k=2\pi/\lambda$ is the wavevector of the probe laser. \Fref{fig:fig4}(b) shows the electric field dependence of $\chi_{\rm i}$ for both positive and negative scan directions, revealing an enhancement in the suppression obtained for the positive scan data where the anti-blockade states are excited first. Unlike the $S$-states, which are all Stark-shifted to lower energy in the presence of ions, the different sign and magnitudes of the Stark shift for the different $m_j$ levels of $D$-state causes ionization to give a similar suppression signature to the cooperative non-linearity observed in \fref{fig:fig3}. Without an independent measurement of the ion fraction present in the cloud it is difficult to separate these two competing mechanisms, however from the loss observed at higher densities the enhanced suppression of the positive scan is most likely due to ionization of the short-range pairs causing an irreversible ion blockade to dominate over the coherent excitation blockade since the Coulomb potential has longer range.

An interesting feature of \fref{fig:fig4}(b) is that in addition to enhanced suppression, there is a transition from a $\chi^{(2)}$ non-linearity for the negative scan direction where the cooperative effect is expected to dominate, to a $\chi^{(3)}$ non-linearity for the positive data where ions appear to play a major role. Both non-linear coefficients show a quadratic density scaling consistent with many body effects as shown for the $\chi^{(3)}$ data in \fref{fig:fig4}(c), with non-linearities of $\chi^{(2)}=5\times10^{-6}$~mV$^{-1}$ and $\chi^{(3)}=5\times10^{-7}$~m$^2$V$^{-2}$ measured at the maximum density. Comparison with the results of Hau \etal{} \cite{hau99} who demonstrated the largest atomic Kerr non-linearity to date, reveal we have achieved a comparable $\chi^{(3)}$ but at two orders of magnitude lower densities.

As noted in the introduction, in standard optical media observation of a $\chi^{(2)}$ scaling requires a crystalline structure that is not symmetric under inversion. It is therefore an unusual result for an atomic system, and further work is needed to explore the role of dynamics on the optical response in the regime of attractive interactions.

\subsection{Resonant dipole--dipole interactions}\label{sec:microwave}

Another interesting property of Rydberg atom interactions is that in addition to tuning the sign and magnitude of the interaction strength via choice of orbital angular momentum state and $n$, it is possible to change the scaling of the interactions from the van der Waals $1/R^6$ potential to $1/R^3$ resonant dipole--dipole interactions. This is achieved by tuning the energies of two Rydberg pair states into resonance with each other, either through an electric field tuning (known as a F\"orster resonance  \cite{gall94}) or via resonant coupling to a close-lying state using a microwave field.

\begin{figure}[t!]
\centering
	\includegraphics[width=11cm]{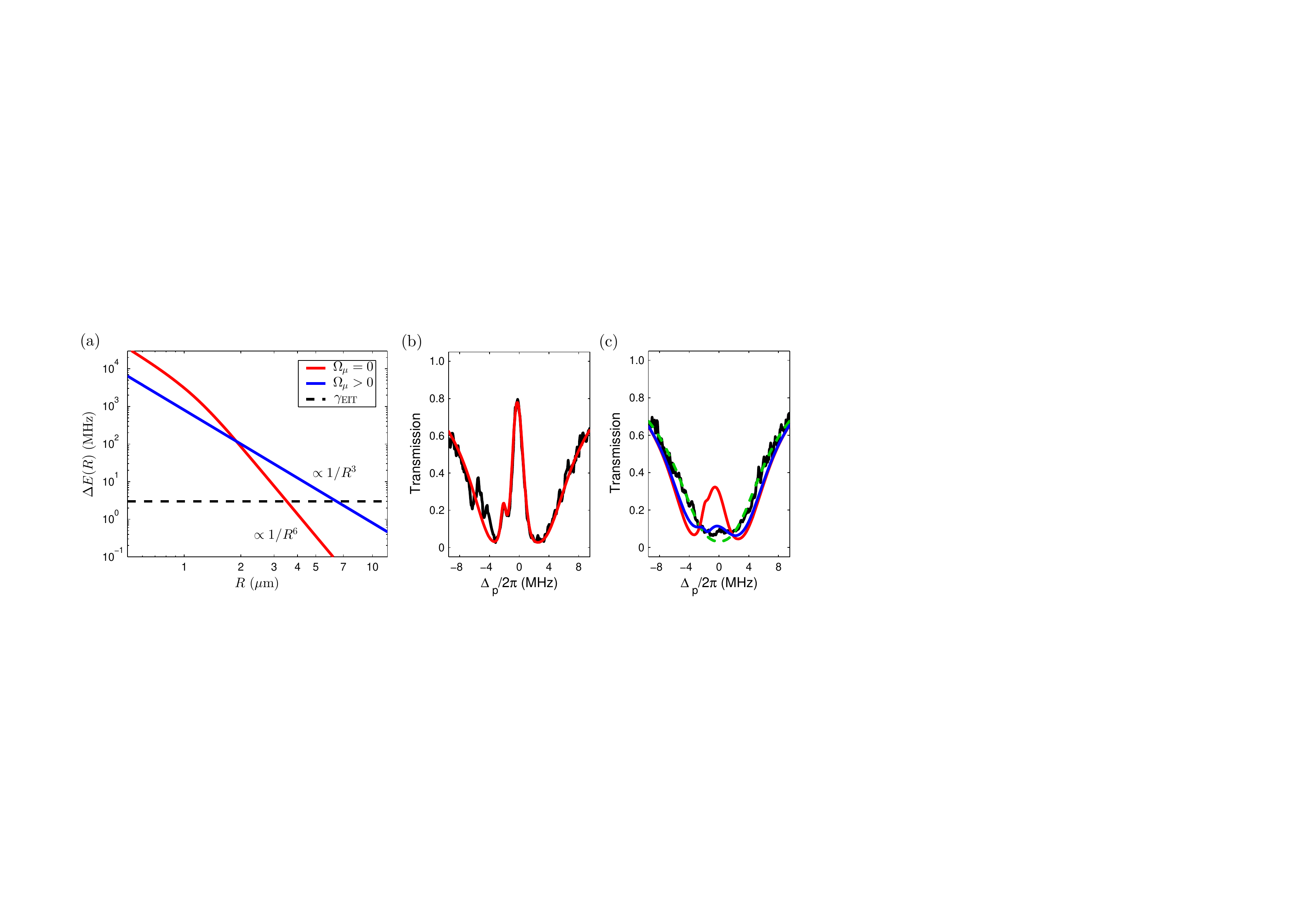}
	\caption{\label{fig:fig5} Enhanced suppression of EIT for weak microwave dressing. (a) Interatomic separation dependence of the energy shift (expressed in frequency) arising from the interaction between Rydberg states. The microwave dressing changes the 1/$R^6$ dependence to a long range $1/R^3$ interaction, thereby increasing the blockade radius (b) Experimental data (black) and theoretical fit (red) for EIT in the weak probe regime with microwave dressing (c) EIT in the strong probe regime with weak microwave dressing shows complete suppression. The blue (red) curve is the theoretical fit assuming $1/R^3$ ($1/R^6$) interactions, the green curve shows the transmission assuming complete Rydberg blockade.}
\end{figure}

The addition of a resonant microwave coupling to the interacting EIT system was explored by Tanasittikosol {\it et al.} on the transition $5s\rightarrow5p\rightarrow46s\rightarrow45p$ \cite{tana11}. In the weak-probe regime, the microwave field causes an Autler-Townes splitting of the EIT resonance, making it possible to perform microwave spectroscopy of the Rydberg state energies using a non-destructive optical probe. The effect of this resonant coupling on the dipole--dipole interactions is shown in \fref{fig:fig5}(a), where the interaction potentials $V(R)$ are plotted for $46s$ with and without the microwave field compared to the EIT linewidth $\gamma_{\rm EIT}$ that determines the blockade radius (dashed line). This illustrates how the resonant coupling extends the $1/R^3$ interaction to all $R$, leading to an increase in the blockade radius from $3.5~\mu$m to $7~\mu$m, corresponding to a transition from $N_{\rm b}\sim 5$ to 40 from the increase in volume of the blockade sphere. Transmission data are plotted in \fref{fig:fig5}(b) and (c) for weak and strong probe Rabi frequencies respectively.  The data reveal an almost complete suppression of the EIT resonance, in contrast to only a 50\% reduction in transparency obtained without the microwave coupling. Theoretical models assuming a $1/R^6$ (red) and $1/R^3$ (blue) interaction potential clearly illustrate that this suppression is consistent with a change in not only the number of atoms per blockade sphere but also the scaling of the interaction potential. Such microwave manipulation of the nonlinear optics could have many applications such as controlling the interactions between polaritons and detection of atoms in neighboring Rydberg states, and as will be discussed in \sref{sec:bariani}, can be exploited to create single photons from an ensemble with a length scale much larger than the blockade radius.

\subsection{Detection of impurities}\label{sec:impurity}

An alternative application of Rydberg EIT is to use it as a method for the detection of impurities in quantum systems. Such schemes are ideally suited to studies of complex many-body phenomena. Olmos {\it et al.} \cite{olmos11} presented a method where the strong interaction among atoms in Rydberg states can be utilized to amplify the effect of an impurity immersed in an ultracold atomic gas to enable single-shot detection of the impurity atom. The method works by converting the presence of the impurity into a quantum state change in many nearby atoms. In this particular scheme the atoms are transferred to another hyperfine state if an impurity is near, allowing spatially resolved imaging of single impurities. A similar scheme was presented by G{\" u}nter {\it et al,}  \cite{gunter12} who performed numerical simulations of the absorption images expected for Rydberg atoms excited from a quasi-2D gas.

\begin{figure}[h!]
\centering
	\includegraphics[width=8cm]{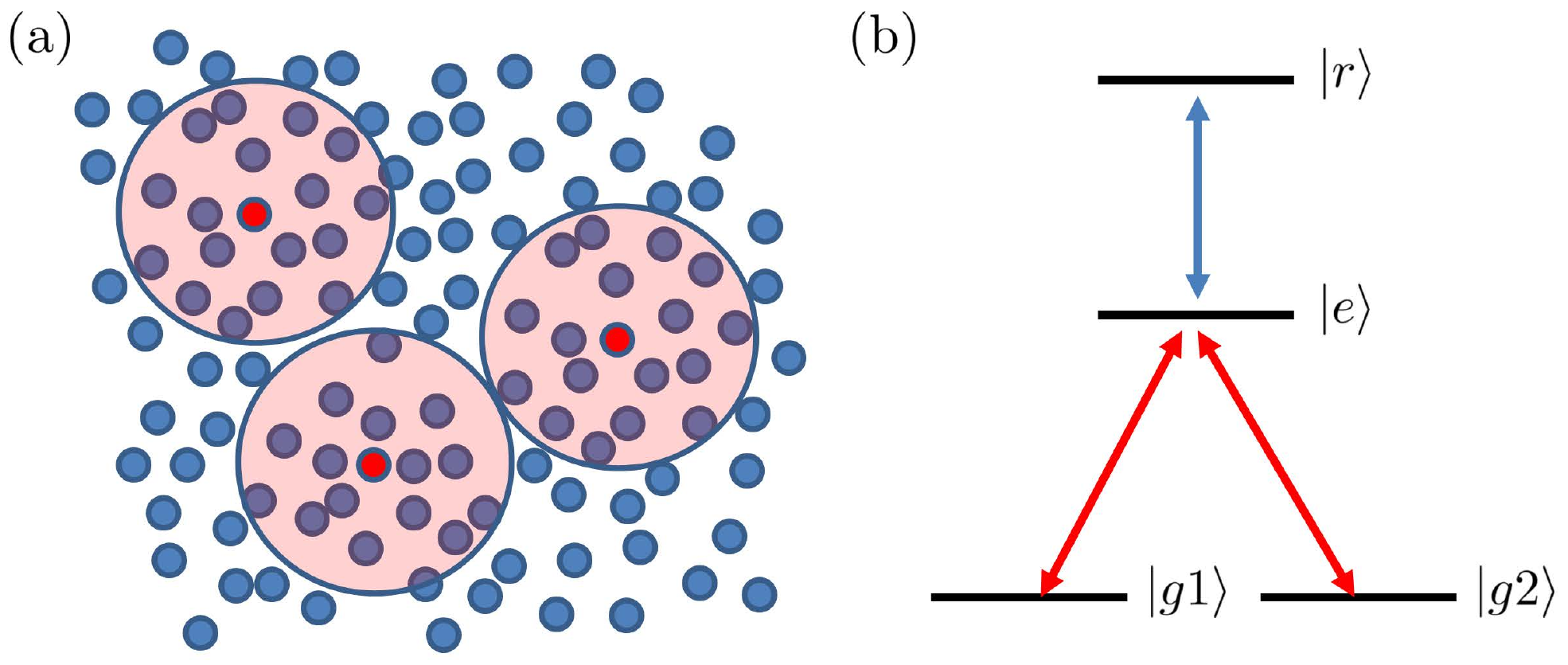}
	\caption{\label{fig:fig5} (a) The presence of impurity atoms (red) causes the neighboring atoms (purple) to change state. (b) Schematic of the energy level scheme. The atoms can be transferred between two ground states $|g1\rangle$ and $|g2\rangle$. Figure adapted from Olmos {\it et al.} Ref.~ \cite{olmo11}.}
\end{figure}

\section{Rydberg Quantum Optics} \label{sec:qo}

In the previous sections the atom-light interactions, and hence non-linearity, have been studied in the semi-classical approximation where the electric field is assumed to be a sinusoidal traveling wave with well defined phase and amplitude. In this basis the cooperative non-linearity can be treated as a large Kerr like non-linearity. Scaling the measured $\chi^{(3)}$ to the BEC densities of $10^{12}$~cm$^{-3}$ and accounting for the quadratic density dependence, the resulting non-linearity is of order $10^{-3}$~m$^2$V$^{-2}$, sufficiently strong to enter the regime of single-photon non-linear optics required for quantum information processing with photons, see Fig.~\ref{fig:fig1}.

At the single photon level however, the semiclassical approximation breaks down and it is necessary to consider the quantum correlations of the light field explicitly. In addition, it has been shown that for even if a conventional optical medium was found with a sufficiently large Kerr effect to induce large phase shifts at the single photon level this would not be sufficient to build an optical quantum computer due to problems such as the self Kerr term distorting pulse propagation \cite{shapiro06,geabanocloche10}.

Dipole blockade provides a novel solution to circumvent this issue as it enables the response of an optical medium to change after absorption of just a single photon, an effect previously only observed under the condition of strong coupling of a single atom to a cavity field \cite{birnbaum05,dayan08}. In this section we will demonstrate how Rydberg atom non-linearities can be exploited to realize photon sources, quantum gates and detectors that form the building blocks required for optical quantum information processing, in addition to developing hybrid optical interfaces to other quantum systems.

\subsection{Photon Sources} \label{sec:sources}

\subsubsection{Deterministic Photon Source}\label{sec:deterministic_single_photon}
\begin{figure}[t!]
	\centering
	\includegraphics[width=9cm,angle=0]{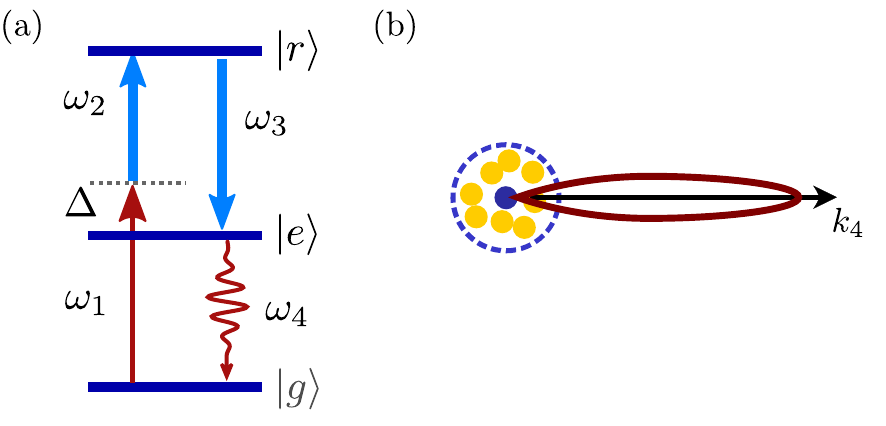}
	\caption{\label{fig:4wmx} Deterministic Photon Source. (a) Four-wave mixing scheme which exploits blockade to deterministically prepare a single excitation in an atomic ensemble. (b) Decay of the singly excited collective state $\ket{\psi}$ results in single photon emission into a narrow cone along the phase-matched direction $\mrb{k}_4=\mrb{k}_1+\mrb{k}_2-\mrb{k}_3$.}
\end{figure}

The original proposal by Saffman and Walker \cite{saffman02} to utilize Rydberg atom dipole blockade for the generation of deterministic single-photons employs the four-wave mixing scheme shown schematically in \fref{fig:4wmx}(a) in an ensemble of atoms confined to a size small compared to the blockade radius $R_{\rm b}$. {The atoms, initially in ground state $\ket{g}$, are subjected to a $\pi$-pulse on the off-resonant two-photon transition at frequencies $\omega_1$ and $\omega_2$ from $\ket{g}$ to $\ket{r}$ with a detuning $\Delta$ from the single photon resonance from $\ket{g}$ to $\ket{e}$ to prevent excitation of the intermediate state.} Dipole blockade prevents excitation of more than a single Rydberg state, leading to {deterministic preparation} of the symmetric collective state $\ket{\psi}=1/\sqrt{\N{}}\sum_j\ket{r_j}$, where $\ket{r_j}$ denotes atom $j$ excited to the Rydberg state and the remaining atoms in $\ket{g}$. A subsequent $\pi$-pulse at frequency $\omega_3$ on the transition from $\ket{r}$ to $\ket{e}$ maps this single excitation onto the superradiant Dicke state
\begin{equation}
\ket{1_{\mrb{k_0}}} = \frac{1}{\sqrt{{\N{}}}}\displaystyle\sum_j^\N{} \e{}^{\ii{} \mrb{k_0} \cdot \bm{r}_j} \ket{e_j}~,
\end{equation}
describing an atomic spin-wave with a single delocalized excitation and wavevector $\mrb{k}_0=\mrb{k}_1+\mrb{k}_2-\mrb{k}_3$. This state will decay to state $\ket{g}$ by the cooperative emission of a single photon into mode $\mrb{k}_4$ with an angular emission probability given by \cite{saffman02}
\begin{equation}
P(\mrb{k}_4) \propto \frac{1}{\N{}}\left|\displaystyle\sum_j\e^{-\ii(\mrb{k}_4-\mrb{k}_0)\cdot\mrb{r}_j}\right|^2~,
\end{equation}
resulting in a strongly directional emission in a narrow cone along the phase-matched wave-vector $\mrb{k}_4=\mrb{k}_0$, even for small atom numbers as illustrated schematically in \fref{fig:4wmx}(b).

A complete calculation of the temporal dynamics of the decay of state $\ket{1_{\mrb{k_0}}}$ from an array of $\N{}\sim1000$ Rb atoms confined within an $8~\mu$m volume reveals a superradiant enhancement in the decay rate proportional to $N_{\rm b}$ at short times  \cite{mazets07,pedersen09}. This is due to the highly correlated collective dipole shared equally across the atoms. During the superradiant phase there is a 95\% probability that photons are emitted with divergence angle $< 0.3~$mrad along $\mrb{k}_4$, {which is robust to loss of atoms from random locations within the array as the collective excitation is delocalised within the ensemble}. Angular emission profiles for alternative geometries have also been investigated  \cite{porras08}.

Deterministic preparation of single-photons via creation of the collective state therefore offers a number of advantages over free-space coupling to single atoms, namely a highly directional emission that can be efficiently collected and a $\sqrt{\N{}}$-enhancement enabling fast readout times $\sim\mu$s  \cite{saffman05}, whilst negating the requirement for single atom addressability.

Preliminary characterization of the phase-matching condition for cw four-wave mixing using a Rydberg state was performed by Brekke \etal{} \cite{brek08} in a non-interacting regime, showing efficiencies of up to 50\% for scattering light in the forward direction. This work has recently been extended to a blockaded ensemble by Dudin and Kuzmich  \cite{dudin12}, who demonstrated a single-photon source with a 2.5~$\mu$s repetition rate when probing a 15~$\mu$m long sample coupled to the $n$s Rydberg states. The single-photon character of the resulting emission can be quantified from the normalized intensity correlation function $g^{(2)}(\tau)$, with a perfect single photon source giving $g^{(2)}(0)=0$ and a coherent state $g^{(2)}(0)=1$. The authors show a smooth variation from $g^{(2)}(0)=1$ for the 40s state to $g^{(2)}(0)=0.04$ at 102s, a clear signature of single photon emission with an efficiency of 10\%. This result is significant not only as a first demonstration of Rydberg atom non-linearities at the single photon level, but also as a state-of-the-art high-brightness single photon source that produces photons with higher efficiency and smaller $g^{(2)}(0)$ at a rate three orders of magnitude faster than the best source of Zhao \etal{} \cite{zhao09} based on the DLCZ protocol \cite{duan01} who achieved a 1~ms duty cycle.

It is possible to further extend this single-photon creation scheme by using the concept of collective encoding \cite{brion07}, where qubit states are coded into the collective populations of the hyperfine states of an atomic ensemble. Nielsen and M\o{}lmer \cite{nielsen10} show this combination enables deterministic preparation of arbitrary entangled photon states such as GHZ, Bell or cluster states between photons in different modes using a single blockade volume. If atoms are instead confined in a ring lattice, Rydberg interactions can be used to prepare cylindrically symmetric collective states, leading to the emission of hollow single photons or spatially correlated entangled photon pairs  \cite{olmos10}.

\subsubsection{Atom-light interfaces}

Studies of single-photon creation and emission also provide information about the efficiency and fidelity of single-photon coupling to a blockaded atomic ensemble, since the process of collective absorption of single photons can be considered simply as the time-reversal of the collective emission detailed above. An efficient coupling between ensembles is important as it allows long-distance transfer of quantum information, and can also be utilized for high fidelity quantum memories or single photon detection using the amplification schemes presented in \sref{sec:impurity}.

Numerical simulations by Pedersen and M\o{}lmer  \cite{pedersen09} demonstrate that the spatio-temporal profile of a single photon emitted from one atomic ensemble can be efficiently absorbed by a second spatially separated ensemble with a fidelity of order 95\%. Combining this with the ability to perform two-atom quantum gates within a collectively encoded ensemble \cite{brion07}, an entanglement pumping algorithm \cite{jiang07} can be utilized to generate entanglement between two ensembles with very high fidelity, thus realizing scalable quantum networks with robust interfaces between flying and stationary qubits \cite{pedersen09}.  Saffman \etal{} \cite{saffman08} propose Holmium atoms as the ideal candidate for implementation of such collective encoding schemes, as the rich internal structure offers up to 60 uniquely addressable qubit states that provide sufficient overhead to permit a practical error-corrected realisation of this protocol.

{\subsubsection{Probabilistic Photon Sources }\label{sec:bariani}

A key experimental challenge associated with using small atomic ensembles to generate single photons is to confine a sufficient enough atoms within a blockade radius of $\sim10~\mu$m to achieve directional emission from the collective state, enabling creation of high fidelity sources. An alternative solution proposed by Bariani \textit{et al.} \cite{bariani12} and Stanojevic \etal{} \cite{stanojevic12} that circumvents the requirement for localization to a single blockade sphere is to prepare a weak coherent excitation from $\ket{g}$ to $\ket{r}$ of a cloud width $\sigma>R_\mr{b}$, resulting in a Poissonian probability distribution of the number Rydberg atoms within the cloud. The strong dipole--dipole interactions between neighbouring atoms can then be exploited to induce an inhomogeneous dephasing of the multiply excited spin-waves during a time $T$, whilst leaving the singly excited states unaffected. Following a subsequent read-out pulse at frequency $\omega_3$ the singly excited state remains phase-matched along $\mrb{k}_4$, with a larger cloud size resulting in enhanced collimation of the emitted photon. For the multiply excited states, the inhomogeneous dephasing destroys the coherence of the imprinted spin-wave which suppresses emission along this axis, with photons emitted in arbitrary directions.

In the Van der Waals regime, Stanojevic \etal{} \cite{stanojevic12} show that for an average interatomic separation $R$, the relevant timescale for dephasing is $T_R=R^6/\vert C_6\vert$, making it possible to create non-Gaussian states with non-classical correlation functions on timescales of order 700~ns. For long interaction times ($T\gg1$), the result is a mapping from a coherent state $\ket{\alpha}$ to the zero or one photon states $\ket{0}+\alpha\ket{1}$ in the phase-matched output mode \cite{stanojevic12}, producing single photons with a probability of $\vert\alpha\vert^2/(1+\vert\alpha\vert^2)$.

In the work of Bariani \textit{et al.} \cite{bariani12} the effect of dephasing due to a resonant microwave coupling from an $n$s to $n'$p Rydberg state is considered, which switches the interactions to long-range resonant dipole-dipole interactions $\propto1/R^3$, as discussed above in \sref{sec:microwave}. The dephasing is controlled using a microwave pulse sequence capable of achieving significant single-photon character after a single cycle of a few microseconds in duration, with an exponential reduction in the second order correlation function obtained through application of multiple dephasing cycles. Using a second microwave coupled state, this scheme can be extended to generate entanglement in the atomic sample and hence mapped onto entangled photon pairs.

The limitation of both of these dephasing schemes is that the timescale for dephasing due to interactions must be short compared to the coherence time of the singly excited spin wave. This is typically limited by motional dephasing caused by atoms moving on a length scale comparable to the wavelength of the spin-wave, $\lambda'=2\pi/(k_1+k_2)$, necessitating use of cold atoms for the required microsecond timescales. This constraint can be relaxed using a four-level Doppler-free excitation scheme, resulting in $\lambda'\rightarrow\infty$ \cite{bariani12}.

\begin{figure}[b!]
	\centering	
	\includegraphics{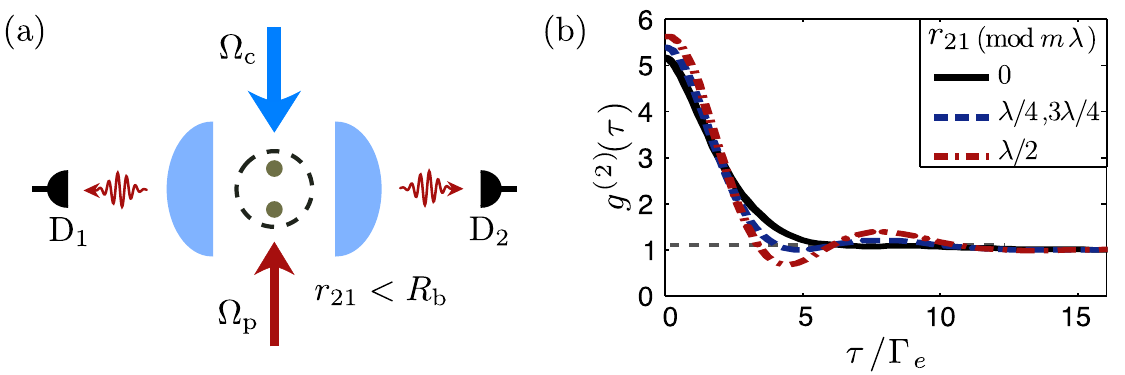}
	\caption{\label{fig:photon_pair} Photon Pair Source. (a) Two atoms confined within a blockade radius scatter light cooperatively due to dipole--dipole interactions, with scattered light being collected using high numerical aperture lenses and a pair of detectors. (b) Analysis of the second-order correlation function for the detectors in (a) reveals strong bunching corresponding to correlated emission of photon pairs that is robust to the inter-atomic separation, $\bm{r}_{21}$.}
\end{figure}

The photon sources considered so far rely on controlled preparation of a large atomic ensemble, however the strong interactions can also be observed through studying the off-axis correlations of the scattered light from a few-atom ensemble. The simplest model system in which to explore these correlations is that of two atoms separated by distance $\mrb{r}_{21}<R_{\rm b}$ as illustrated in Fig.~\ref{fig:photon_pair}(a). The atoms are driven under steady-state conditions on the EIT resonance, with scattered light efficiently collected onto a pair of detectors. Calculation of the second order correlation function by Pritchard \etal{} \cite{prit11} reveals a strongly bunched signature associated with the correlated emission of photon pairs, as shown in \fref{fig:photon_pair}(b). This correlation function can be understood from the discussion of the cooperative effect of dipole blockade in EIT presented in \sref{sec:RydEIT}, where the presence of a single Rydberg atom causes the system to resonantly scatter light due to the contribution of the doubly-excited state in the interacting dark state of \eref{eq:drk}. Thus if a photon is detected at $D_1$, the atoms were initially in state $\ket{ee}$ prior to detection, and a second photon is emitted within a few spontaneous lifetimes in the direction of detector $D_2$, causing the observed bunching. This angular correlation in the photon emission arises from the collective nature of the blockade effect, causing a phase-locking of the dipoles conditioned on the first detection event as is observed in superradiance \cite{dicke54}.

For physical implementation as a probabilistic source of photon pairs, it is necessary to consider effects of relative atomic position on the correlation function. \Fref{fig:photon_pair}(b) shows that the bunched correlation function persists as the interatomic separation $r_{21}$ is varied over $\lambda$. This scheme is therefore robust with respect to atomic motion, and taking into account realistic experimental parameters is capable of generating on average one correlated photon pair every 30~$\mu$s \cite{pritchard12}, comparable to the single-photon source demonstrated by Dudin and Kuzmich. \cite{dudin12}

\subsection{From Rydberg blockade to photon blockade}

Quantization of the electric field in vacuum reveals each spatial mode can be described as a quantum harmonic oscillator, resulting in a discrete ladder of energies $\hbar\omega(n+1/2)$ dependent upon the number of photons $n$ within each mode. However, if the harmonicity of this ladder can be broken, it is possible to create non-classical states of light from a classical coherent state input $\ket{\alpha}$. An example of this is the interaction between a single atom and the optical field inside a cavity as illustrated in \fref{fig:anharmonic}(a). The atom--field coupling results in a splitting of the harmonic ladder into states $\ket{n,\pm}$ with a detuning of $\pm\sqrt{n}g_0$, where $\pm$ denotes the ground and excited state of the atom and $g_0$ is the coupling strength. In the strong coupling regime, where the splitting is larger than the cavity linewidth, the cavity is shifted out of resonance with the probe beam after the first photon enters the cavity, preventing subsequent photons from entering the cavity until the first photon leaves. Thus, by introducing anharmonicity after the first excitation, it is possible to create a single-photon filter. This effect is known as photon blockade \cite{imamoglu97}. The resulting anti-bunched output has been observed experimentally for both a single atom in an optical cavity  \cite{birnbaum05,dayan08} and, more recently, in superconducting cavities coupled to a Cooper pair box  \cite{lang11}. Achieving the required strong-coupling regime with a single atom in the cavity field is experimentally challenging, however exploiting the long-range interactions of the Rydberg states opens the possibility of observing photon blockade without the requirement of an optical cavity.

\begin{figure}[t!]
	\centering
	\includegraphics{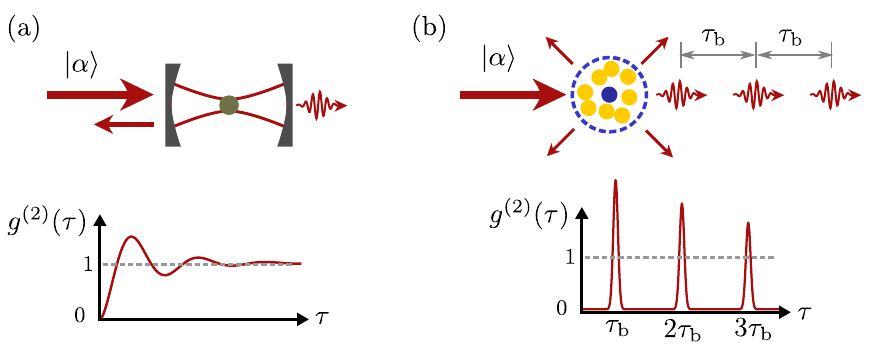}
	\caption{\label{fig:anharmonic} Photon Blockade. (a) A single atom in a cavity detunes the cavity from resonance after a single photon is absorbed, preventing more than one photon to pass through the cavity to create a single-photon output from a coherent input pulse. (b) An optically thick ensemble of atoms confined within a blockade radius can create a photon blockade as only a single photon can form a dark state polariton, leading to a train of single photons being transmitted with characteristic delay time $\tau_{\rm b}$ for the polariton to propagate through the ensemble.}
\end{figure}

In the strongly interacting Rydberg system, anharmonicity at the single photon level is provided by the dipole blockade mechanism, which shifts the system out of resonance with the optical field after absorption of a single photon, analogous to the atom--cavity system. For the case of Rydberg EIT, the first photon can form a dark state and is transmitted without loss through the blockade region. Subsequent photons are now detuned from the EIT resonance due to the strong dipole--dipole interactions, causing them to be scattered resonantly out of the probe beam mode, as illustrated in \fref{fig:anharmonic}(b). To understand the properties of the resulting photon-blockaded output, it is necessary to consider the optical propagation of a single photon through an EIT medium.

For a 3--level system, the Maxwell-Bloch propagation equations can be solved analytically to reveal that the combined light-matter excitation forms a stable, loss-less quasiparticle known as a dark state polariton \cite{fleischhauer00} that can be written as
\begin{eqnarray}
\psi(z,t)&=&\cos\theta \epsilon_{\rm p}(z,t) -\sqrt{N}\sin\theta \rho_{\rm gr}(r,t){\rm e}^{{\rm i}(k_{\rm c}-k_{\rm p})z}~,
\end{eqnarray}
where $\epsilon_{\rm p}={\cal E}_{\rm p}/\sqrt{\hbar\omega_{\rm p}/2\epsilon_0}$ is the normalized probe field and $\rho_\mr{gr}(r,t)$ describes an atomic spin-wave excitation that is phase-matched to the probe and coupling lasers with wavevectors $k_\mr{p}$ and $k_\mr{c}$ respectively. The mixing angle, which controls the weighting of the superposition between an excitation of the electromagnetic field and atomic spin-wave, is related to the group index $n_{\rm g}$ by
\begin{eqnarray}
\tan^2\theta&=&n_{\rm g}~\approx\chi_0\frac{\omega_{\rm p}\Gamma}{\Omega_{\rm c}^2}.
\end{eqnarray}
For large group index the input field is nearly completely converted into an extended atomic excitation inside the medium, making it possible to `store' and then retrieve a photon by varying the control field $\Omega_{\rm c}$, which is the basis for quantum memory. \cite{hamm10}.

In order to map the Rydberg blockade onto a photon blockade we want to localize the dark state polariton inside the blockade sphere. The length of the photon wavepacket outside the medium is $\ell=c/\Delta \nu$ where $\Delta \nu$ is the bandwidth of the photon pulse. If we match the bandwith to the EIT bandwidth ($\Delta_{\rm EIT}\sim \Omega_{\rm c}^2/2\Gamma$) \cite{fleischhauer05} then the length of the pulse inside the medium, i.e. the length of the dark state polariton, is
\begin{eqnarray}
\ell_{\rm dsp}&=&\frac{\ell}{n_{\rm g}}=\frac{c/\Delta_{\rm EIT}}{6\pi N \omega_{\rm p}/2k_{\rm p}^3\Delta_{\rm EIT}}=\frac{2k_{\rm p}^2}{6\pi N}~.
\end{eqnarray}
Using the fact that the optical depth of a medium with length equal to $R_{\rm b}$ is
\begin{eqnarray}
{\rm OD}&=&N\sigma R_{\rm b}=\frac{6\pi N R_{\rm b}}{k^2_{\rm p} }~,
\end{eqnarray}
where $\sigma$ is the resonant two-level scattering cross-section of each atom, we can write that
\begin{eqnarray}
\ell_{\rm dsp}&=&\frac{2 R_{\rm b}}{\rm OD}~.
\end{eqnarray}
Thus, providing the optical depth is greater than 2, we can localize the excitation inside a single blockade sphere.

As the superatom can only support one Rydberg excitation, only one Rydberg polariton can pass through at a time. Any other photons that arrive during the transit time of the Rydberg polariton $\tau_{\rm b}=R_{\rm b}/v_{\rm g}$ are scattered. A classical input field with a random distribution of photon times is converted into a highly non-classical photon pulse train with a pulse spacing of $\tau_{\rm b}$. This regular pulse train is evidence of strong single photon interactions. The resulting correlation function is illustrated schematically in \fref{fig:anharmonic}(b), which shows well resolved bunched peaks at integer multiples $\tau_{\rm b}$. This is analogous to the correlation function for atoms localized on a 1D lattice, showing the strong Rydberg atom interactions can be mapped onto the optical field to create hard-sphere photons. A fundamental difference between this correlation function and that illustrated in (a) is that in the cavity system the characteristic width of the anti-bunching time is related to the mean number of photons in the initial coherent state $\ket{\alpha}$ \cite{lang11}, whilst the single-photon separation $\tau_{\rm b}$ for the blockaded ensemble is independent of photon number \cite{gorshkov11}. This effect can therefore be exploited to create a deterministic single-photon source providing there is sufficient optical depth within a blockade radius to achieve both a group delay larger than the incident pulse-width and efficiently scatter the higher-order photons out of the mode. An alternative scheme to generate a similar strongly correlated photon train with well defined temporal separation is through mapping many-body crystalline excitations of Rydberg atoms in a 1D trap onto an optical field  \cite{pohl10}.}

The pioneering experiments on Rydberg EIT \cite{prit10} operated at low density where the OD per blockade sphere was much less than unity and consequently the photon blockade effect would be difficult to observe. However, the regime of photon blockade is well with reach of current experimental techniques are results are expected very soon.

Guerlin \etal{} \cite{guerlin10} have considered theoretically the combined regimes of \fref{fig:anharmonic} for the case that a Rydberg ensemble is placed inside an optical cavity. Dipole blockade enables the ensemble to behave as an effective two-level superatom, making it possible to exploit the collective $\sqrt{\N{}}$ scaling to reach the strong-coupling regime for efficient simulation of the Jaynes Cummings Hamiltonian \cite{jaynes63}. As well as simplifying the process of obtaining photon blockade in the cavity, the collective coupling results in a larger splitting of the cavity modes, making multi-photon resonances clearly resolvable to produce controlled bunching of light output from the cavity.

\subsection{Quantum Gates} 

\begin{figure}[t!]
\centering{}
	\includegraphics{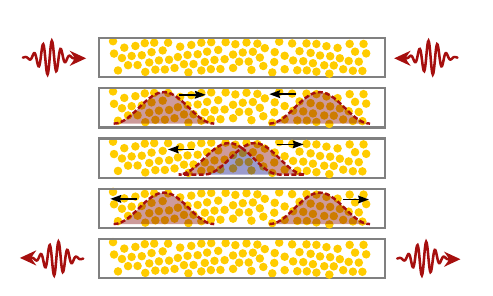}
	\caption{\label{fig:gates} Photonic Phase Gate. Two counter propagating photons enter a long, optically thick medium and are converted into dark-state polaritons which propagate through each other. As the polaritons overlap, strong dipole--dipole interactions prevent multiple Rydberg excitations within radius $R_{\rm b}$, destroying the polariton eigenstate and converting the excitation back into a photon which propagates at velocity $c$ through the overlap. The result is a conditional phase-shift in the event that two photons are present in the medium simultaneously, which can be tuned to $\pi$ permitting implementation of a photonic quantum gate.}
\end{figure}

Rydberg atom photon blockade describes the behavior on the resonant EIT condition, where the medium is transparent for a single photon and opaque for higher-order photons which are scattered from the blockade region. Off-resonance however ($\Delta_{\rm p}=-\Delta_{\rm c}=\Delta$), the dipole--dipole interactions can be exploited to create a conditional phase-shift between photons, as illustrated in \fref{fig:gates}. Two counter-propagating photons enter an optically thick atomic ensemble and are each converted into dark-state polaritons, each with a single Rydberg excitation. As the two polariton wave-functions overlap, the strong dipole--dipole interactions prevent both Rydberg excitations being localized within a blockade radius, and the polaritons are no longer eigenstates of the system, causing the photons to travel at velocity $c$ through the blockaded overlap region.

Analytic expressions from Gorshkov \etal{} \cite{gorshkov11} show that the two-photon wavefunction thus acquires a factor of $\exp(-\ii\varphi-\eta)$, where $\varphi \sim -\Gamma/4\Delta d_{\rm B}$ is the phase-shift and $\eta \sim\Gamma^2/8\Delta^2 d_{\rm B}$ is the absorption (loss), and $d_{\rm B}$ represents the optical depth across the blockade radius. For $d_{\rm B}\gg1$ and $\Delta\gg\Gamma$, an almost loss-less phase shift of $\pi$ can be realized to implement a conditional phase-gate for two photons, with the phase-shift only acquired if both photons are present in the medium.

Friedler \etal{} \cite{friedler05} show a conditional phase gate can also be obtained by coupling the incident photons to different Rydberg states by using independent control fields. Rather than exploiting the blockade effect, an external field is applied to polarize the Rydberg states and induce resonant dipole--dipole interactions, leading to an accumulated phase-shift as the two polaritons pass through each other in the medium for Rydberg states with large dipole moments. This concept was extended by Shahmoon \etal{} \cite{shahmoon11} to consider 1D propagation in a hollow-core fibre.  In addition to studying the quantum phase gate, the authors also propose a method to achieve a quantum non-demolition (QND) measurement of photon number when a weak coherent state $\ket{\alpha}$ is coupled to a low-lying Rydberg state with small dipole moment, and a photon number state $\ket{n}$ is coupled to a state with large dipole moment. Following propagation through the system, the coherent state acquires a phase proportional to the number of photons $n$, which can be measured using homodyne detection.

For application as a phase-gate it is important to provide strong confinement of the transverse modes to ensure a spatially uniform phase-shift across the photon wavepacket, as for counter-propagating photons with a finite transverse overlap the conditional phase-shift can cause diffraction, compromising gate fidelity  \cite{he11}. This can be circumvented by using cold atoms loaded into a hollow-core photonic crystal waveguide \cite{shahmoon11,gorshkov11} to reduce the optical field to 1D propagation whilst achieving optical depths $d_{\rm B}>1$ required for a loss-less phase-gate.

The other option to overcome the issue of transverse modes is to use spatially separated optical modes. An example of this approach is to consider using two atomic ensembles placed within the microwave field of a superconducting stripline cavity  \cite{petrosyan08}, which couples non-resonantly to the Rydberg atoms in each ensemble to extend the range of the dipole--dipole interaction over several millimeters, as discussed in \sref{sec:hybrid}.

\subsection{Detection}

The final element required for efficient quantum information processing with photons is a high fidelity measurement of the output state. Generally this is achieved using single-photon avalanche photodiodes or photomultiplier tubes that can discriminate between 0 photons or at least 1 photon arriving at the detector within a given time window. Photon number resolution can be obtained with these devices either through spatial \cite{leaf07} or temporal \cite{achilles04} multiplexing, or as recently demonstrated by using a novel self-differencing circuit to obtain 0-4 photon resolution \cite{kardynal08}. However, these devices typically suffer from low quantum efficiencies, limiting the detection fidelity.

Rydberg atoms offer a number of solutions to this issue, where the enhanced collective coupling due to the strong interactions can be exploited to realize high efficiency conversion from photonic to atomic excitations. An example of this was presented in \sref{sec:deterministic_single_photon}, where the single-photon coupling between two spatially separated blockaded spheres resulted in a transfer efficiency in excess of 95\% \cite{pedersen09} due to the emitted photon having an optimal spatio-temporal profile for absorption.

\begin{figure}[!t]
	\centering
	\includegraphics{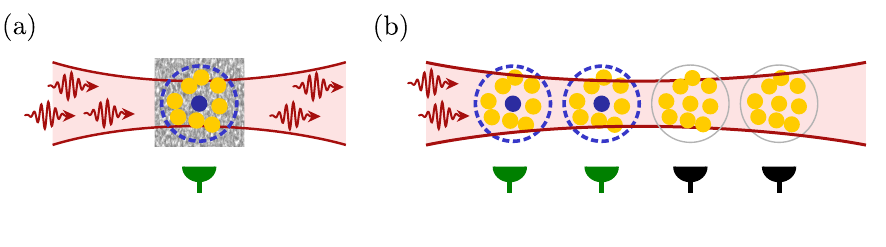}
	\caption{Photon Detection. (a) Photon subtraction is achieved by superimposing a speckle pattern on a blockaded ensemble. Dephasing due to the uncorrelated AC Stark shifts of the Rydberg states cause the photonic excitation to be trapped in the atomic medium, deterministically subtracting a single photon from the probe beam which can then be detected. (b) An array of successive photon subtractions provides photon number resolving detection. Figures adapted from Ref.~ \cite{honer11}.\label{fig:detector}}
\end{figure}

This dependence on the efficiency with pulse-shape can be relaxed using a refinement to the collective absorption scheme proposed by Honer \etal{} \cite{honer11} which achieves deterministic and untriggered single photon subtraction from a probe beam, as illustrated in \fref{fig:detector}(a).  A weak few-photon probe laser drives an off-resonant two-photon transition from $\ket{g}$ to $\ket{r}$. Assuming $R<R_\mathrm{b}$ the ensemble is excited to state $\ket{\phi}=1/\sqrt{\N{}}\sum\ket{r_j}$, removing a photon from the probe pulse, which will then be emitted cooperatively back into the probe mode. However, if an inhomogeneous broadening of the Rydberg states is introduced that causes the phase of the Rydberg state of each atom to evolve at a different rate, the collective state is dephased, with population shared out into the $\N{}-1$ many-body dark-states that are not dipole coupled to the ground state $\ket{g}$, trapping the single-photon excitation as a Rydberg excitation in the medium. This is similar to the microwave dephasing discussed above, however it is all states rather than just pair-states that are dephased. Realization of such a dephasing mechanism is possible by imaging a laser speckle pattern with short-range spatial and temporal correlations into the ensemble that creates a local time-dependent AC Stark-shift on the Rydberg state of each atom. Deterministic single-photon subtraction is achieved providing the dephasing occurs faster than the propagation time of the photon through the medium, without stringent conditions on the pulse shape. This also offers a novel scheme to generate non-classical light states, such as a Schr\"{o}dinger cat-state from squeezed vacuum.

Both of these protocols offer an extremely efficient transfer of excitation from a single photon to a single Rydberg atom, which can be ionized and detected using a channel electron multiplier with greater than 98\% efficiency \cite{henkel10}. Alternatively, the state of the incident photonic qubit can be measured optically by weakly probing the transmission or dispersion, or through the impurity amplification schemes described in \sref{sec:impurity}. With all of these techniques, the optical transmission of a weak probe with many photons is controlled by the presence of the incident photon, analogous to creating a single-photon transistor  \cite{honer11}. Finally, photon number resolution is achieved using an array of ensembles and recording the number of ensembles containing an excitation as illustrated in \fref{fig:detector}(b) and, unlike the number resolving photodiode schemes mentioned above, each photon can be detected with equal probability.

\subsection{Experimental Progress}
Experimental progress towards observing these single-photon non-linearities requires development of an efficient scheme to simultaneously confine both ground and Rydberg states in an atomic ensemble with sufficient optical depth. Since magnetic traps for Rydberg atoms have been found to be effective only for high angular momentum states  \cite{lesa05}, optical dipole traps or lattices are being explored. However, optical traps have a number of associated problems, for instance, the large polarizability of the Rydberg states causes a significant position dependent AC Stark shift of the Rydberg states within the trap, leading to inhomogeneous broadening and reducing the fidelity of the blockade effect. Optical traps are also repulsive for Rydberg atoms due to the ponderomotive potential experienced by the Rydberg electron. To circumvent these issues several avenues have been explored. For example, the phase modulation of a red detuned optical lattice produces a time averaged potential which traps the Rydberg atoms along the longitudinal trap direction  \cite{ande11}. Another solution is to use blue-detuned `hollow' or `bottle beam' traps where both the ground and excited states are attracted to regions of low intensity  \cite{isen09}. In such traps the AC stark shifts are minimized because the atoms are located in regions of low field. Zhang {\it et al.} showed that for careful choice of trapping laser wavelength, trap geometry and the addition of a background compensation field, a `magic' trapping condition is achieved where ground-state and Rydberg atoms experience the same level shifts at the center of the trap  \cite{zhan11}. In all optical trapping schemes the photoionization rates should be considered as they can be significantly less than the radiative decay rates of the Rydberg states  \cite{adams06}. Magic wavelength trapping schemes for Rydberg atoms also have promise for in-situ thermometry for metrology applications, since Rydberg states are very sensitive blackbody shifts  \cite{ovsi11}.

An alternative solution to free-space trapping is to utilize the strong transverse confinement of a hollow-core photonic crystal waveguide, as required for the proposed photon--photon phase-gates. These have been utilized in a number of experiments using both atomic-vapor and laser-cooled atom clouds to exploit the large optical depth resulting from a mode-matching between the probe beam and absorption cross-section, for example to realize an optical switch triggered by a pulse of only a few hundred photons \cite{bajcsy09}. One of the challenges of implementing Rydberg gates in such a system is the proximity of the atoms to the fibre wall, as Rydberg atoms strongly couple to both the mirror image in the dielectric and to surface polariton modes, as observed using Rydberg EIT in a thin quartz microcell \cite{kubler10}. However, this can be overcome by careful choice of principal quantum number to find a state that is not resonantly coupled to the surface modes.

\subsection{Hybrid Quantum Optical Interfaces}\label{sec:hybrid}

Hybrid quantum devices combine a number of different physical qubit realizations to exploit the optimum properties of each, for example long storage times in one and fast gates in another. It is also possible to couple systems of discrete qubits to those with continuous quantum variables, such as position and momentum, to offer further computational enhancement  \cite{lloyd00}. Circuit quantum electrodynamics (QED)  \cite{blais04} offers such an architecture, where micro-fabricated superconducting stripline cavities with frequencies in the microwave regime are coupled to Cooper pair boxes, acting as artificial atoms with the qubit basis realized in the charge degree of freedom. These qubits provide very fast gate operations but suffer from rapid decoherence rates  \cite{wendin05}, and whilst a chip-based realization is intrinsically scalable, flying qubits are difficult to implement.

\begin{figure}[!t]
	\centering
	\includegraphics{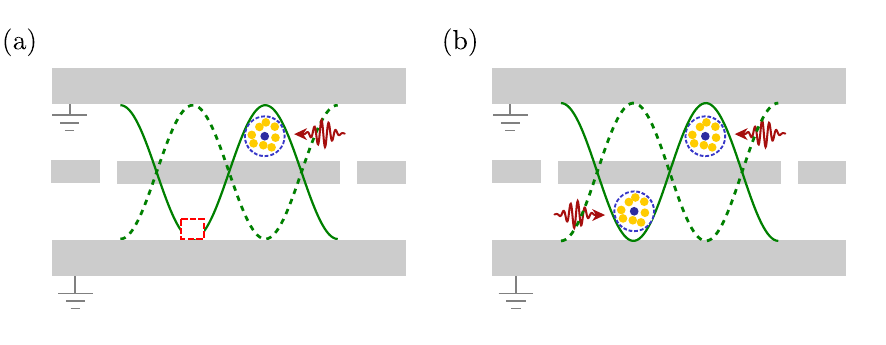}
	\caption{Hybrid Quantum Interfaces. Resonant coupling between the microwave field of a superconducting stripline cavities and transitions between high-lying Rydberg states can be exploited to (a) couple Rydberg atoms to a superconducting qubit located at the anti-nodes of the cavity field and (b) achieve long range non-local coupling between photons addressing different Rydberg ensembles. Figures adapted from Refs.~ \cite{petrosyan08,petrosyan09}.\label{fig:stripline}}
	\end{figure}

Exploiting the strong coupling of the Rydberg state to external electronic and magnetic fields, as well as the microwave-transition frequencies between close lying states, it is possible to create a strong coupling to other quantum architectures, such as circuit QED. For example, Rydberg atoms held above a superconducting wire experience a capacitive coupling to the cavity mode \cite{sorensen04}, enabling long-distance interactions between atoms separated by millimeters at either end of the wire, or coupling to a superconducting qubit. Instead, an ensemble can be placed within the microwave field of a strip-line cavity tuned close to a near-resonance in the Rydberg state  \cite{petrosyan09}, with the solid-state qubits interacting with the ensemble through the cavity mode as illustrated in \fref{fig:stripline}(a). Thus the fast solid-state gates can be used for computation and the atomic ensemble for storage. Combining the atom-qubit coupling via the cavity with the collective single photon emission and absorption discussed above, Rydberg states therefore offer the potential to create quantum optical interfaces for circuit QED to enable optical read-out of the solid state qubits or alternatively deterministic detection of single photons. An analogous hybrid interface between a solid state qubit and a single ion has recently been suggested \cite{kielpinski12}, however the collective nature of the ensemble Rydberg coupling provides the benefit of directional emission over the isotropic scattering of a single ion.

Petrosyan \etal{} \cite{petrosyan08} also demonstrate how these stripline microwave cavities can also be utilized to generate long range non-local coupling between photons addressing different atomic ensembles, as illustrated in \fref{fig:stripline}(b). The resulting photon--photon interactions are sufficiently strong to realize a quantum phase gate, whilst the separation of the spatial modes of the photons remedies the issue of transverse mode diffraction for single mode entanglement \cite{he11}.

Experimental realization of such a hybrid quantum device is challenging due to the strong optical fields required to localize an atomic ensemble close to the surface of a superconductor, which if scattered on the surface cause a local heating that destroys the resonant quality of the cavity, resulting in decoherence. Progress towards this goal has been achieved by observation of a coherent interaction between a beam of Helium atoms and a co-planer microwave waveguide cooled to 100 K  \cite{hogan12}. A possible alternative is to use a piezoelectric nano-mechanical resonator in place of the strip-line cavity which can capacitively couple to both a superconducting qubit and a Rydberg atom  \cite{gao11}. This has the advantage of being smaller than a strip-line cavity for enhanced scalability, and can be placed directly above the superconducting qubit to shield the effects of scattered light.

\section{Outlook}

Rydberg states provide an ideal resource for non-linear optical media, with a diverse range of possible applications from electrometry to quantum information. The desired properties can be optimized by choice of principal quantum number and angular momentum state, or tuned by application of external electric or magnetic fields.  These properties can be efficiently mapped onto strong optical transitions, allowing non-destructive read-out for classical applications like electrometry or large single photon non-linearities  for quantum information applications.

The strong long-range dipole--dipole interactions of the Rydberg states enable cooperative behavior to be observed for atoms separated by more than an optical wavelength, creating a novel `non-local' optical non-linearity. Dipole blockade leads to a collective enhancement of the coupling between a single photon and an atomic ensemble by a factor~$\sqrt{N_{\rm b}}$ that removes the requirement of single-atom addressability and is robust to atom loss. This enables deterministic and directional single photon sources suitable for use as high-fidelity quantum interfaces between stationary and flying qubits in a quantum network. Dipole blockade also provides a mechanism to achieve strong photon--photon interactions. In the resonant regime this is manifest as a photon blockade creating a correlated train of single photons with well defined temporal separation; in the off-resonant regime it can create a conditional phase-shift for applications as a two-photon quantum gate. As well as coupling photons to atoms, the Rydberg states open the possibility to create hybrid quantum optical interfaces to solid-state qubits to create novel photonic quantum processors.

The topic of Rydberg non-linear optics is at a very early stage and the most exciting developments and breakthroughs are yet to come. Hopefully this brief review provides a taste of why there are reasons to be excited and gives some pointers about where to start.

\bibliography{Rydberg_Review_Article}

\end{document}